\documentclass[twocolumn]{aastex631}
\usepackage{amsmath}
\usepackage{mathrsfs}

\begin{document}

\title{Monte Carlo evaluations of gamma-ray and radio pulsar populations}
\correspondingauthor{Shawaiz Tabassum} 

\author[0000-0002-4222-5027]{Shawaiz Tabassum}
\affiliation{Department of Physics and Astronomy, West Virginia University \\
P.O. Box 6315, Morgantown, WV 26506, USA}
\email{st00034@mix.wvu.edu}
\affiliation{Center for Gravitational Waves and Cosmology, Chestnut Ridge Building, Morgantown, WV 26505, USA}

\author[0000-0003-1301-966X]{Duncan R. Lorimer}
\affiliation{Department of Physics and Astronomy, West Virginia University \\
P.O. Box 6315, Morgantown, WV 26506, USA}
\affiliation{Center for Gravitational Waves and Cosmology, Chestnut Ridge Building, Morgantown, WV 26505, USA}
\email{Duncan.Lorimer@mail.wvu.edu}

\begin{abstract}
Based on well-grounded Galactic neutron star populations formed from radio pulsar population syntheses of canonical pulsars (CPs) and millisecond pulsars (MSPs), we use the latest Fermi-LAT catalog (4FGL-DR4) to investigate the implications of proposed $\gamma-$ray luminosity models. Using Monte Carlo techniques, we calculate the number of CPs and MSPs that would comprise the sample of pulsar-like unidentified sources (PLUIDs) in 4FGL-DR4. While radio beaming fractions were used to scale the sizes of the populations, when forming the mock 4FGL-DR4 samples, we make the simplifying assumption that all $\gamma-$ray pulsars are beaming towards the Earth. We then explore the observable outcomes of seven different $\gamma-$ray luminosity models. Four of the models provide a good match to the observed number of PLUIDs, while three others significantly over-predict the number of PLUIDs. For these latter models, either the average beaming fraction of $\gamma-$ray pulsars is more like 25--50\%, or a revision in the luminosity scaling is required. Most of the radio detectable MSPs that our models predict as part of the PLUIDs within 4FGL-DR4 are, unsurprisingly, fainter than the currently observed sample and at larger dispersion measures. For CPs, in spite of an excellent match to the observed radio population, none of the $\gamma-$ray models we investigated could replicate the observed sample of 150 $\gamma-$ray CPs. Further work is required to understand this discrepancy. For both MSPs and CPs, we provide encouraging forecasts for targeted radio searches of PLUIDs from 4FGL-DR4 to elucidate the issues raised in this study. 
\end{abstract}

\section{Introduction} \label{sec:intro}

Pulsars are immensely useful astrophysical probes for studying the underlying nuclear, plasma and gravitational physics that power a wide variety of observed phenomena \citep[for recent reviews see, e.g.,][]{2016arXiv160207738K,2022hxga.book...30N}. Originally discovered in the radio band \citep{1968Natur.217..709H}, they have since been shown to emit rotation-powered pulses across the electromagnetic spectrum \citep[see Chapter 9 in][and references therein]{2012hpa..book.....L}. Among the many high-energy instruments so far, Fermi Large Area Telescope (Fermi-LAT) has been revolutionary in opening the $\gamma-$ray window for pulsars \citep{2009ApJ...697.1071A}. Since its launch in 2008, Fermi-LAT has enabled the discovery of 294 pulsars that emit in $\gamma-$rays, making them 10\% of all known pulsars 
\citet[][hereafter 3PC]{2023ApJ...958..191S}. { About a quarter of these Fermi-LAT discovered pulsars are so far only observable in $\gamma-$rays.} These sources are also the dominant Galactic component of $\gamma-$rays above 1~GeV. Furthermore, $\gamma-$ray luminosity of some pulsars has been observed to account for up to all of their total available spin-down energy (See Figure 24 in 3PC). This signifies the importance of studying $\gamma-$ray emission to better understand the high energy physics of pulsars. Understanding their Galactic population, such as their spatial distribution, total number and the number detectable by our telescopes, is important not only for stellar evolution studies but also to help us better constrain pulsar emission mechanisms. Population synthesis can be a viable tool for this purpose and will form the basis of the present study. 

Pulsars broadly fall into two categories: millisecond pulsars (MSPs), with spin periods $P \lesssim 30$~ms, and canonical pulsars (CPs), with $P \gtrsim 30$~ms \citep[for a recent review, see, e.g.,][]{2022ASSL..465....1B}. While accounting for the differences in period derivative, $\dot{P}$, between these two distinct populations, this simple cut-off in period separates the population into the older, recycled MSPs and the young CPs. Radio pulsar population syntheses have been used to constrain properties{,} such as the birth rate and the period distribution{,} for {the Milky Way's underlying} CP and MSP population in multiple studies over the past half century \citep[see, e.g.,][]{1970ApJ...160..979G, 1985MNRAS.213..613L, 1990ApJ...352..222N, 1997ApJ...482..971C, 2006ApJ...643..332F, 2014MNRAS.443.1891G, 2015MNRAS.450.2185L, 2020MNRAS.492.4043C}. Among other things, these studies have highlighted the difficulties in being able to constrain these distributions due to a range of different models being able to explain the observed population. Multi-wavelength population syntheses are one avenue to help remove these degeneracies, while also probing emission models at other wavelengths. 

For CPs, \cite{2004ApJ...604..775G} implemented such a scheme by incorporating $\gamma-$ray pulsars which had been discovered by the Energetic Gamma$-$Ray Experiment Telescope (EGRET), the predecessor to Fermi-LAT \citep{1995ApJS..101..259T, 2001AIPC..558..103T}. This allowed \cite{2004ApJ...604..775G} to test the slot-gap model of $\gamma-$ray emission geometry \citep{2003ApJ...588..430M} and also conclude that the radio emission model of \cite{2002ApJ...568..289A} was insufficient to explain the observed population. They were limited by the number of known radio-loud $\gamma-$ray pulsars at the time, which was the seven detected by EGRET. After the launch of Fermi-LAT, various studies \citep{2011ApJ...726...44T, 2011ApJ...727..123W, 2012A&A...545A..42P} were able to improve on the earlier works due to the increase in the number of $\gamma-$ray CPs and better understand the underlying radio population. These authors were able to test different models of $\gamma-$ray emission geometries but were still limited by the small sample ($\sim$ 25) of known radio loud $\gamma-$ray CPs. 

More recently, \cite{2020MNRAS.497.1957J} studied the highest energy and youngest CPs by imposing a cutoff on spin-down energy of $\dot E > 10^{35}$ erg s$^{-1}$. They assumed an empirical $\gamma-$ray luminosity law, as seen among Fermi-LAT CPs, of $L_{\gamma} \propto \sqrt{\dot E}$, while using the outer-gap (OG) emission geometry \citep{1986ApJ...300..500C}. All of these works assumed a spin-down model of a vacuum retarded-dipole as first introduced by \cite{1955AnAp...18....1D}. \cite{2022A&A...667A..82D} investigated all CPs and also used a better spin-down model based on the work of \cite{2014ApJ...781...46C}, which accounts for charged particles surrounding the pulsar magnetosphere in a force-free regime. They were able to test the empirical two dimensional (2D) plane luminosity relation for $\gamma-$ray luminosity proposed by \cite{2019ApJ...883L...4K}, which includes a weak power-law dependence on the magnetic field, along with the $\sqrt{\dot E}$ term. The geometry used was based on the striped wind model \citep{2011MNRAS.412.1870P, 2021A&A...654A.106P}. {\cite{2024A&A...691A.349S} built upon the work of \cite{2022A&A...667A..82D} by including a Galactic potential model in which the pulsars evolve over time and a minimum dispersion measure in their $\gamma-$ray detection model while still using the 2D plane luminosity relation and a binary Fermi-LAT sensitivity cutoff based on Galactic latitude. Their results showed that the striped wind model is able to reproduce the distribution of peak-separation seen in the $\gamma-$ray light curves of the observed $\gamma$-ray CP sample.}
 
For MSPs, \citet{2007ApJ...671..713S} carried out a pioneering study before the launch of Fermi-LAT that aimed to test the polar cap model for $\gamma-$ray emission geometry. \citet[][hereafter GHF+18]{2018ApJ...863..199G} applied it to the radio and $\gamma-$ray emitting MSP population and derived $\gamma-$ray luminosity laws based on the emission geometries predicted by three emission models: the OG, the slot gap two-pole caustic \citep[TPC;][]{2003ApJ...598.1201D} and the pair-starved polar cap  \citep[PSPC;][]{2005ApJ...622..531H}. They were also able to implement a more realistic spin-down model with a magnetosphere in a force-free regime. In another effort to extract the luminosity law for Fermi-LAT MSPs, \cite{2018MNRAS.481.3966B} carried out a fully Bayesian analysis to constrain parameters for four different statistical distributions. They found a broken power-law to be preferred over all other luminosity functions that they tested. Similarly, \cite{2019ApJ...883L...4K} showed that the $\gamma-$ray luminosity of all Fermi-LAT pulsars, CPs and MSPs, lies on a plane in log-space. This fundamental plane (FP), parameterized either by just the magnetic field strength and the spin-down energy in the 2D case or by including spectral cut-off energy as well to form the three dimensional (3D) plane, was improved using a larger sample  of $\gamma-$ray pulsars in \citep{2022ApJ...934...65K}. This relationship was shown to be consistent with the one theoretically expected if the observed $\gamma-$ray emission is the result of curvature radiation. 

All of these studies made use of models where emission takes place within the light cylinder of the pulsar, at gaps in the magnetosphere where charge densities are sufficiently low to allow particle acceleration to take place along magnetic field lines. This particle acceleration is hypothesized to produce the observed $\gamma-$rays through curvature, synchrotron or inverse Compton radiation. However, recent work that has been able to model the magnetosphere under force-free conditions (i.e., the absence of plasma pressure and the condition $ \vec{E} \cdot \vec{B} = 0$, where $\vec{E}$ is the electric field and $\vec{B}$ is the magnetic field) have shown that the observed $\gamma-$ray emission by Fermi-LAT originates in an equatorial current sheet, mainly beyond the light cylinder in a region known as the separatix layer\footnote{Named as such because it represents a region which separates the open and closed magnetic field-lines.} \citep{2010ApJ...715.1282B, 2010MNRAS.404..767C, 2014ApJ...793...97K}. In \cite{2023ApJ...954..204K}, it was shown that the FP luminosity relation is also consistent with a model in which $\gamma-$ray emission occurs due to curvature radiation in the region around the separatix layer known as the separatix zone. This was achieved using a 3D kinetic global magnetosphere in the force-free regime, further substantiating the results and consequently also the FP relation. \cite{2020JCAP...12..035P} was able to independently check the consistency of FP with the Fermi-LAT data using a $\gamma-$ray pulsar population synthesis. To the best of our knowledge, the statistical relations, specifically the 3D FP, have so far not been tested in a multi-wavelength population synthesis. We stress here that multi-wavelength population synthesis does not mean a synthesis which separately treats the population of pulsars in multiple wavelengths. Instead, it implies that the single synthesized population is applied to surveys at multiple wavelengths at the same time.

Motivated by the goal of further understanding the nature of the unassociated Fermi-LAT sources, we approach the problem using population syntheses that are informed by our knowledge of the radio CP and MSP populations. We start by synthesizing a population consistent with the observed radio population of CPs and MSPs by accounting for selection effects from major blind radio surveys using PsrPopPy \citep{2014MNRAS.439.2893B}. We then test the effectiveness of multiple different $\gamma-$ray luminosity models using the sample of known $\gamma-$ray pulsars. Our aim is to investigate the efficacy of various $\gamma-$ray luminosity laws from the works of \cite{2018MNRAS.481.3966B}, \cite{2018ApJ...863..199G} and \cite{2023ApJ...954..204K} {in being able to produce a sample of pulsars compatible with the observed sample. Our work is not intended to constrain the best model. Instead, we elucidate the ways in which these models succeed and fail at reproducing the observed sample of pulsars after all the known multi-wavelength selection effects have been accounted for. We do select a fiducial $\gamma-$ray luminosity model to investigate more parameters and} place limits on the number of $\gamma-$ray pulsars waiting to be discovered among the unassociated Fermi-LAT $\gamma-$ray sources. We finally assess the limitations of this work and suggest possible avenues for improvement in the future.

\section{Fermi Catalog and $\gamma-$ray Pulsars} \label{sec:catalog}

The latest Fermi catalog at the time of this work, 4FGL-DR4, contains 7194 sources \citep{2022ApJS..260...53A, 2023arXiv230712546B}. While the majority of these are blazars, the dominant source from our Galaxy is pulsars. Out of the 294 known $\gamma-$ray pulsars, 150 are CPs and 144 are MSPs. At the time of writing, {of the observed $\gamma-$ray pulsars,} 70 CPs are radio-quiet\footnote{As defined in 3PC, they have radio flux density at 1400 MHz of at most 30~$\mu$Jy.} whereas only 6 MSPs are radio-quiet. We exclude two of the radio-loud MSPs that are in globular clusters (J1824-2452A \& J1823-3021A) from our analysis. We note that, while the double neutron star binary PSR~J0737--3039A is also (by our $P<30$~ms definition) deemed a $\gamma-$ray MSP \citep{2013ApJ...768..169G}, its different evolutionary history means that we also excluded it from our sample of 141 $\gamma-$ray MSPs. For these 141 MSPs, 7 lack a reported $\dot P$ value in the catalog and are also excluded from our sample when comparing $P- \dot P$ plots or $\dot P$ distribution. These $\dot P$ lacking MSPs do contribute when comparing sizes of the samples. For CPs, we use the complete sample of 150. There are also 2428 sources in the catalog that remain unassociated with any astrophysical object and hence unidentified as an astronomical source. Out of these unidentified sources, 1061 are most likely to be pulsars due to the constraints on their observed properties. These constraints require the sources to have a log-parabolic spectrum, a spectral cut-off energy of 4~GeV and their fluxes do not significantly vary. Variability index \citep[see section 3.6 of][]{2012ApJS..199...31N} is a proxy for the $\gamma-$ray flux variability from the mean value, with a value of greater than 24.7 indicating that there is a less than 1\% chance of it being a steady source \cite[for details, see][]{2022ApJS..260...53A}. Trends seen in the observed $\gamma-$ray pulsar population guide these constraints. \citep{2023ApJ...958..191S}. A population synthesis of $\gamma-$ray pulsars should not predict the number of Fermi-LAT detectable $\gamma-$ray pulsars to be greater than the number of pulsar-like unidentified sources (henceforth referred to as PLUIDs) plus the number of known $\gamma-$ray pulsars. 

\begin{figure}[ht!]
\plotone{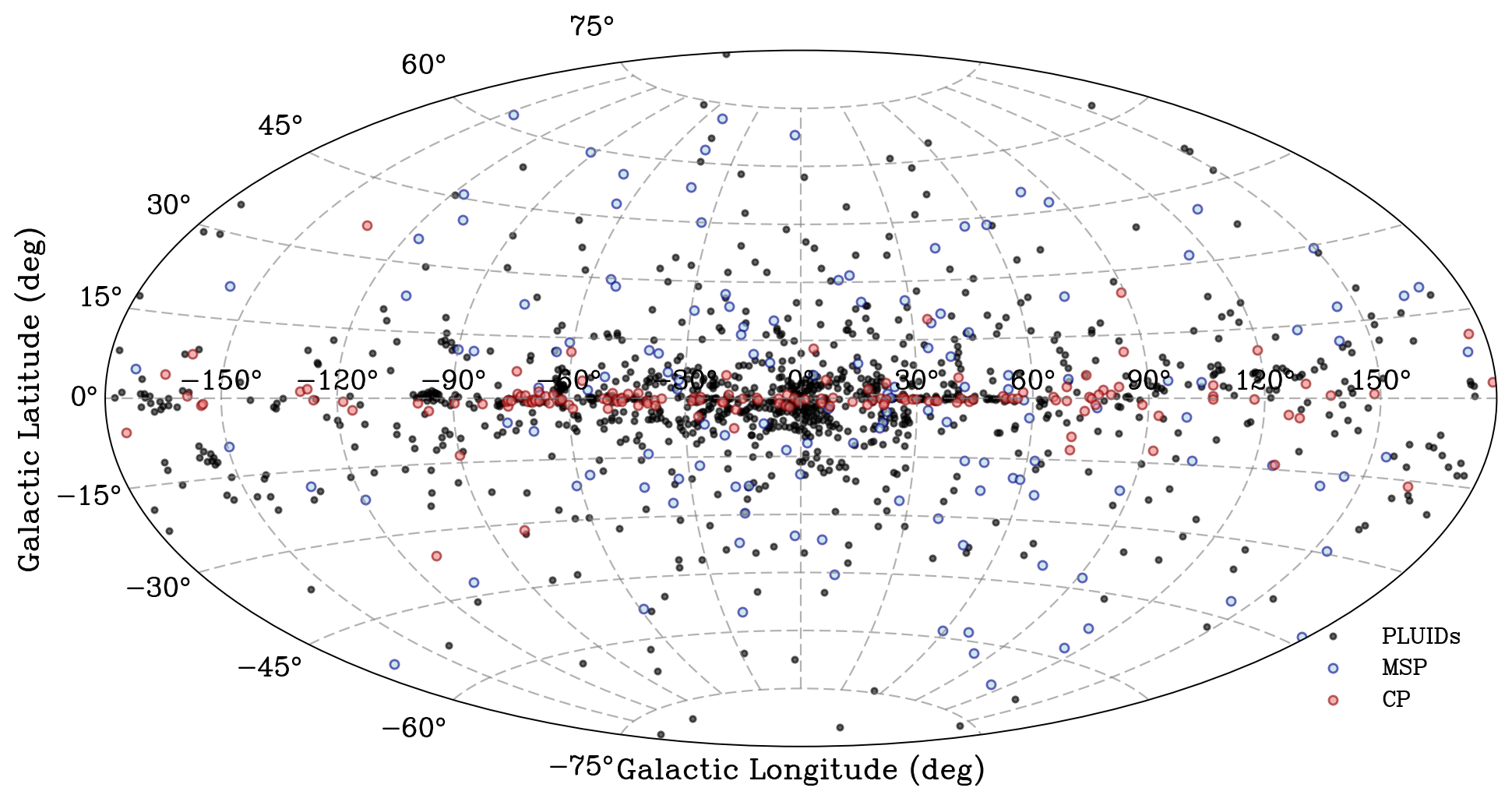}
\caption{Distribution in Galactic coordinates showing $\gamma-$ray pulsars and PLUIDs in the 4FGL-DR4 Fermi-LAT catalog.
\label{fig:aitoff-uids}}
\end{figure}

In Fig.~\ref{fig:aitoff-uids} we show an Aitoff projection of the spatial distribution of the PLUIDs and the known $\gamma-$ray pulsars in the Galaxy. It is difficult to distinguish if a PLUID is a MSP or a CP based on its spectral properties alone. As such, without making any assumptions, we can only come up with upper limits by considering all PLUIDs to be either CPs or MSPs. This means that a population model should not predict more than 1203 MSPs (142 Galactic MSPs + 1061 PLUIDs) or 1211 CPs (150 Galactic CPs + 1061 PLUIDs) in our Galaxy. Similarly, it should not predict less than 150 CPs and 142 MSPs. We can refine these limits further for the MSPs by assuming only PLUIDs with Galactic latitudes $|b|>10^\circ$ could be MSPs and even from those, assume only half are MSPs. This is a reasonable assumption since MSPs are old pulsars which have been moving in our Galaxy's gravitational potential for long enough to have moved away from the galactic plane. We see evidence for this by looking at the known $\gamma-$ray MSPs distribution on the Aitoff plot in Fig. \ref{fig:aitoff-uids}. This assumption gives us a lower limit of 312 (142 known Galactic MSPs plus 340/2 PLUIDs with $|b|>10^\circ$). Similarly, for the upper limit we can assume that all the  $|b|> 10^\circ$ PLUIDs are MSPs. We can further assume that in the $|b|< 10^\circ$ range, the proportion of MSPs to CPs is the same in PLUIDs as is in the observed sample which is 23\%. This gives us an upper limit of 648 for MSPs (142 known Galactic MSPs plus 340 PLUIDs with $|b|>10^\circ$ plus (0.23 $\times$ 721 of PLUIDs with $|b|< 10^\circ$)). For CPs, the upper limit is 705, resulting from an assumption that the rest of the 77\% $|b|<10^\circ$ PLUIDs are CPs. The lower limit remains 150 since it is harder to untangle the $|b| < 10^\circ$ sources which most likely contain a large number of other non-pulsar astrophysical sources such as pulsar wind nebulae or supernova remnants. We use these limits to confront our model pulsar populations in Section~\ref{sec:results}.

\section{Synthesizing Pulsars} \label{sec:style}

Using a modified version of the pulsar population synthesis code PsrPopPy{\footnote{https://github.com/shawaizt/PsrPopPy-G}} \citep{2014MNRAS.439.2893B}, we perform Monte Carlo simulations of pulsars observable by a set of radio surveys {listed in Table \ref{tab:1}}. We start by obtaining a birth period, age, magnetic field and magnetic inclination angle from appropriate distributions{, as listed in Table \ref{tab:2}}. Each pulsar is then placed in the model galaxy by first finding its radial position from the model galaxy's center and then using this to place it in one of the spiral arms, following the procedure outlined in Appendix B of \cite{2006ApJ...643..332F} and using updated parameters for the spiral arms as given in \cite{2017ApJ...835...29Y}. The pulsar's $z$-component of spatial position is obtained through sampling a double-sided exponential distribution with a particular scale height. Next, each model pulsar is evolved in a model of the Galactic potential and its present day position in the Galactic plane is obtained. We used the adaption of the \citet{1987AJ.....94..666C} model of the Galactic potential given by \citet{1989MNRAS.239..571K}. 

Each pulsar is spun down according to a specific model and its present day period and period derivative are recorded. At this point a model may be applied to see whether it is beaming towards Earth. If it is not, the pulsar is discarded and the previous steps are restarted until we obtain a pulsar which is beaming towards us. Then its pulse width is calculated, followed by its radio luminosity and flux at Earth. The flux is calculated using the geometric distance assuming that the sun is 8.5~kpc from Earth. We then calculate its dispersion measure based on its Galactic location and distance, using the YMW16 electron density model \citep{2017ApJ...835...29Y}. Before the final step, all the parameters of the pulsar are recorded. Finally, we check which surveys would have observed the location of this pulsar in the sky, and if its flux would be above the sensitivity limit of that survey at that location. If it is detectable by at least one of the surveys, a counter is incremented. This process is repeated until the counter is equal to the total number of pulsars detected by all of the surveys being used for the synthesis. Details about the distributions and models used specific to MSPs and CPs are given in Table \ref{tab:2} and explained below.

The steps outlined above are fairly germane to population syntheses of radio pulsars. In this work, we develop a multi-wavelength population synthesis 
where we use the Galactic neutron star populations 
of CPs and MSPs as inferred from radio surveys and subsequently apply a set of filters to generate samples of Fermi-LAT detectable $\gamma-$ray pulsars.
The $\gamma-$ray selection effects are applied to the total Galactic population of pulsars obtained from the synthesis using a sensitivity map of the Fermi-LAT 4FGL-DR3 catalog  \citep{2022ApJS..260...53A}. This is different to the single flux cut off applied by \citet[]{2022A&A...667A..82D}. The resulting number of sources represent the total number of pulsars detectable by Fermi-LAT as point sources. Hence this number should be equal to the total number of Fermi-LAT MSPs/CPs plus the number of unassociated sources most likely to be pulsars based on their point source properties.

\subsection{Observed Population and Radio Surveys}

We use the Parkes Multi-beam Survey (PMSURV), Parkes-Swinburne Multi-beam Intermediate (SWINIL) and High latitude (SWINHL) surveys for our synthesis of CPs. These surveys have detected a total of 1323 unique CPs in the Galactic plane according to the ATNF Catalog\footnote{http://www.atnf.csiro.au/research/pulsar/psrcat} \citep{2005AJ....129.1993M}.  For MSPs, we used the 92 unique MSPs that have been detected in the seven surveys listed in Table \ref{tab:1}. It is to be noted that the total of each column in Table \ref{tab:1} is greater than the aforementioned number of total pulsars in each survey since some pulsars are detected in multiple surveys. These surveys were selected because they have a wide sky coverage, were conducted at L-band and had a uniform observing system for the duration of the surveys. Data from these have also been thoroughly searched using different techniques \citep[for a discussion, see e.g,][]{2023MNRAS.522.1071S}. The sample size of CPs from those three surveys accounts for 46$\%$ of all known Galactic CPs. The number of MSPs are also half of all known MSPs discovered from blind surveys i.e., excluding MSPs discovered through targeted searches of Fermi$-$LAT sources\footnote{https://www.astro.umd.edu/$\sim$eferrara/pulsars/GalacticMSPs.txt}. Using PsrPopPy, we {set up our model using} the following survey parameters: antenna gain, integration time at each pointing, sampling time, system temperature, center frequency, bandwidth, channel bandwidth, number of polarizations, full-width half maximum of the beam, signal-to-noise ratio cutoff used by the survey, minimum and maximum Galactic longitude and latitudes observed, fraction of the specified Galactic region that has been observed and a survey degradation factor to account for the effects of finite sample \citep{2012hpa..book.....L}. Further details can be found in \citet{2014MNRAS.439.2893B}.

\subsection{Modeling considerations for millisecond pulsars}

Most of the birth parameters for MSPs that we used are from the work of \citet[][hereafter LEM+15]{2015MNRAS.450.2185L}. With a larger number of MSPs available now compared to LEM+15's study, we decided to use this updated sample to derive a birth period distribution. We define the birth of an MSP as the time when the accretion from its companion stops and it becomes radio loud. Following \citet{2014ApJ...793...97K} and GHF+18, in the ideal force-free magnetosphere, the period as a function of time
\begin{equation} \label{eq:1}
    P(t) = \sqrt{P_{0}^{2} + \frac{2 \pi^{2} R^{6} B_{0}^{2}}{c^{3} I} \bigg(1 + \sin^{2}\alpha \bigg) t},
\end{equation}
where $P_{0}$ is the birth period, $R$ is the radius of the neutron star which we set to 12~km, $B_{0}$ is the birth magnetic field, $c$ is the speed of light, $I$ is the moment of inertia which we set to $1.7\times10^{45}$~g~cm$^{2}$ and $\alpha$ is the magnetic inclination angle. To obtain the birth period ($P_0$) distribution, we first start with the present sample of known $\gamma-$ray MSPs and then evolve them backwards in time using Equation~\ref{eq:1} for the spin evolution to solve for $P_0$. The ages used for the reverse-evolve process were obtained by sampling a uniform distribution from 0 to $10^{9}$ yrs. The magnetic field distribution was assumed to be the same as the one for the forward evolve process (i.e., a log-normal distribution with parameters given in Table \ref{tab:2}).
To compute $\alpha$, we sampled a uniform distribution over a quarter of the unit circle. After these steps, we fit a log-normal distribution to periods of the reverse-evolved population which resulted in a best-fit mean of 0.98~ms with a standard deviation of 0.52~ms, as given in Table \ref{tab:2}. We note that this is significantly different to a fit of this distribution to the observed population (i.e., $P(t)$), for which we find a mean of 1.3~ms and standard deviation of 0.51~ms. For the subsequent populations of radio MSPs we produce using the forward-evolve process, we apply the beaming model proposed by \cite{1998ApJ...501..270K}.

Numerous $\gamma-$ray luminosity models have been proposed for MSPs using diverse methodologies by multiple researchers. We consider all of those which have been proposed in the past decade. First we have the purely statistical models from \cite{2018MNRAS.481.3966B} which are derived from Bayesian model fitting on the observed sample. Their preferred model is a broken power law (BPL) but the log-normal (LN) distribution also had strong evidence so we test both of those. GHF+18 used the emission geometries from three different vacuum magnetosphere models of MSPs to find exponents for a power law $\gamma-$ray luminosity by doing population synthesis of radio and $\gamma-$ray MSPs. We tested all three of these models and their details are given in Table \ref{tab:3}. Finally, we used the 2D and 3D FP of \cite{2019ApJ...883L...4K, 2022ApJ...934...65K, 2023ApJ...954..204K} since they are also applicable to MSPs. Since the 3D-plane relation requires that the $\gamma-$ray cutoff energy {is} known, we tested two different methods of reliably sampling this. We first tested the quadratic relationship, dependent on $\dot E$, first shown in \cite{2017ApJ...842...80K}, with updated parameters by fitting to the current sample of $\gamma-$ray MSPs. We also fit a Gaussian to the histogram of known cut-off energies, $\epsilon_{\rm cut}$. Doing this in log-space, where $\epsilon_{\rm cut}$ is in units of Mev, we draw $\log_{10} \epsilon_{\rm cut}$ from a normal distribution with a mean of 3.458 and a standard deviation of 0.181. This model statistically agreed better with the observed sample than the quadratic model which is why we used it to obtain cutoff energies for our synthesized pulsars. For the three power-law models from GHF+18 and the fundamental planes, we also applied a noise parameter to account for the statistical uncertainty in measuring fluxes and distances used to estimate the luminosity of the observed sample. This noise component was drawn from a Gaussian with a mean of zero and standard deviation of 0.2 and added to the luminosity in log-space. We ran 10 iterations of this synthesis and then aggregated the results to obtain a mean number of detectable $\gamma-$ray MSPs. This number is presented, along with its standard deviation, in the plots comparing the expected number of detectable MSPs with the number suggested by each model. We note
that our modeling of $\gamma-$ray MSPs does not invoke any beaming models, i.e., we explicitly assume all $\gamma-$ray MSPs are beaming towards Earth. We discuss the implications of this approach below.

\subsection{Modeling considerations for canonical pulsars}

We used a log-normal distribution for the magnetic field and the period. The parameters for the magnetic field were obtained from the work of \citet[][]{2015MNRAS.454..615G}. They show that there are multiple values of the mean and variance for this distribution which are supported by the observed sample. This is why we experimented with multiple of these using our spindown model to find the ones which produced a $P-\dot P$ plot closest to the observed one. The period distribution we used is from the work of \citet[][]{2020MNRAS.497.1957J} who showed this to produce good agreement for energetic CPs. We made the choice of implementing a spin-down model which incorporated an exponential decay in the magnetic inclination angle following the work of \citep[][hereafter JK17]{2017MNRAS.467.3493J}. This model involves solving the ordinary differential equation, 
\begin{equation}  \label{eq:2}
    \dot P = K P^{2-n(t)}.
\end{equation}
The time dependent braking index, $n(t)$, is obtained by first calculating $n^{*}(t)$ using the formulation for a decaying $\alpha$ as introduced by \citet{2001A&A...376..543T} and given by,
\begin{equation} \label{eq:3}
    n^{*}(t) = n_{0} - \frac{3 c^{3} I P^{2}(t) \cos(\alpha(t)) \dot \alpha(t)}{4 \pi^{2} R^{6} B_{0}^{2} \sin^{3}(\alpha(t)) }.
\end{equation}
Finally, to account for stochastic effects on the braking index due to state changes and/or intermittency, $n(t)$ is obtained by sampling a normal distribution with a mean of $n^{*}(t)$ and a standard deviation of $n^{*}(t)/3$. The birth braking index, $n_{0}$, is obtained by sampling from a normal distribution with a mean of 2.8 and standard deviation of 1.0. The form of $\alpha$ decay we use, following JK17, is the simple exponential,
\begin{equation} \label{eq:4}
    \sin(\alpha(t)) = \sin(\alpha_{0}) e^{-t/\tau_{D}},
\end{equation}
where the time constant, $\tau_{D}=$ 5$\times$10$^{7}$ yr. As shown in JK17, this model is sufficient to produce a $P-\dot P$ plot which matches the observed one, without the need for a decaying magnetic field. We verified their results {by looking at the $P-\dot P$ distribution obtained using their simplistic assumption that all pulsars are born with a period of 20~ms and the same $\dot P$. However, we chose to use birth assumptions which are more reasonable, such as a distribution of initial periods and a spin-down model.} This still results in {an} excellent overall match {to} the $P-\dot P$ {distribution} between the CP models and the observed population. We used the beaming model of \cite{1998MNRAS.298..625T}. 

After obtaining a model galaxy seeded with these CPs beaming towards us in radio, we calculate their $\gamma-$ray luminosities using the 2D FP, 3D FP and a heuristic model where $L_{\gamma} \propto \sqrt{\dot E}$. For the heuristic model we used the proportionality constant of $10^{16.5}~$erg~s$^{-1}$, as given in 3PC. These models also had the same noise component mentioned in the previous section added to them. We do not use a $\gamma-$ray beaming model for the 2D and 3D FP relations. For the heuristic model we initially employed a $\gamma-$ray beaming model based on the work of \citet[][]{2024arXiv240611428S}. This beaming model makes use of the $\dot E$ dependent beaming fractions originally highlighted by \cite{2020MNRAS.497.1957J}. However, as discussed further below, we modified it to get {a $P-\dot P$ distribution which is qualitatively most similar to the observed distribution.} 

\begin{deluxetable*}{cccc}[t]
\tablewidth{1.0\linewidth}
\tablecaption{Radio surveys used for this work. The number of detected sources for each class of pulsar are given if the survey was part of that class's synthesis. The references detail the survey parameters but do not necessarily include all the discoveries from the survey. \label{tab:1}
}
\tablecolumns{3}
\tablehead{
\colhead{Survey} & \colhead{$N_{\rm det, MSP}$} & \colhead{$N_{\rm det, CP}$} & \colhead{Reference} 
}
\startdata
Deep Multibeam Survey (DMB) & 2 & - & \cite{2013MNRAS.434..347L} \\
Parkes High Latitude Survey (PHSURV) & 5 & - & \cite{2006MNRAS.368..283B} \\
Perseus Arm Survey (PASURV) & 1 & - & \cite{2013MNRAS.429..579B} \\
Pulsar Arecibo L-band Feed Array (PALFA) & 35 & - & \cite{2006ApJ...637..446C}\\
Swinburne High Latitude (SWINHL) & 8 & 63 & \cite{2009ApJ...699.2009J} \\
Swinburne Intermediate Latitude (SWINIL) & 12 & 158 & \cite{2001MNRAS.326..358E} \\
Parkes Multibeam Survey (PMSURV) & 28 & 1125 & \cite{2001MNRAS.328...17M} \\
\enddata
\end{deluxetable*}

\begin{deluxetable*}{ccc}[t]
\tablewidth{1.0\linewidth}
\tablecaption{Parameters and their values used in the synthesis of the radio population. When a particular distribution is sampled for a given parameter, its name is given followed by its mean and standard deviation in parentheses. For the case of the uniform distribution, the numbers in parentheses represent the interval. The truncated normal distribution is truncated to be always greater than zero. \label{tab:2}
}
\tablecolumns{3}
\tablehead{
\colhead{Parameter} & \colhead{MSP} & \colhead{CP} 
}
\startdata
Age (yrs) & Uniform$(0, 5\times 10^{9}$) & Uniform$(0, 10^{8})$ \\
Initial period (ms) & Log-normal(0.98, 0.52) & Normal(50, 10) \\
Spin-down model & Eq.~\ref{eq:1} & Eqs.~\ref{eq:2}, \ref{eq:3} \& \ref{eq:4} \\
log$_{10}$ [Magnetic field (G)] & Normal(8.2, 0.3) & Normal(12.7, 0.3) \\
Radial distribution (kpc) & Truncated Normal(0, 8.5)) & Truncated normal(7.04, 1.83) \\
Scale height (kpc) & Exponential with mean 200~pc & Exponential with mean 50~pc \\
Birth velocity (km/s) & Normal(0, 70) & Normal(0, 180) \\
Pulse width & Eq. 5 in \cite{2006MNRAS.372..777L} & Eq. 5 in \cite{2006MNRAS.372..777L} \\
Radio luminosity & Eq. 14 in \cite{2010MNRAS.404.1081R} & Eq. 18 in \cite{2006ApJ...643..332F}  \\
\enddata
\end{deluxetable*}

\begin{deluxetable*}{ccc}[t]
\tablewidth{1.0\linewidth}
\tablecaption{All the $\gamma-$ray luminosity models, their functional form and best fit parameters that were used in this work. The units for luminosity for all models are erg s$^{-1}$ except for the GHF+18 models; OG, PSPC \& TPC, for which they are eV s$^{-1}$. For the FP models, B should be in units of Gauss, $\dot E$ should be erg s$^{-1}$ and $\epsilon_{\rm cut}$ should be MeV. In the GHF+18 models, $P$ is in units of ms and $\dot P$ is in units of $10^{-21}$s/s. $\dot E$ is in units of erg s$^{-1}$ for the heuristic model. The range for luminosity in BPL and LN is [$10^{30}, 10^{37}$]. The FP, GHF+18 and heuristic models have a noise component drawn from a zero-mean Gaussian distribution with standard deviation 0.2 added to them in log space.  \label{tab:3}
}
\tablecolumns{3}
\tablehead{
\colhead{Model} & \colhead{Functional Form} & \colhead{Parameters} 
}
\startdata
BPL & $dN/dL \propto \begin{cases} 
                                        L^{-\alpha_{1}} & L\leq L_{b} \\
                                         L^{-\alpha_{2}} & L_{b}< L  
                                    \end{cases}
                    $ & $\alpha_{1}=0.97$, $\alpha_{2}=2.60$, $L_{b}=10^{33.24}$  \\
LN & $dN/dL \propto \frac{1}{L} \exp{[-\frac{(\log_{10}L - \log_{10}L_{0})^{2}}{2\sigma_{L}^2}]}$  & $\sigma_{L}=0.63$, $L_{0}=10^{32.61}$ \\
2D FP & $L_{\gamma} = K B^{a} \dot E^{b}$ &  $K=10^{15.0}$, $a=0.11$, $b=0.51$  \\
3D FP & $L_{\gamma} = K B^{a} \dot E^{b} \epsilon_{\rm cut}^{c}$ & $K=10^{14.3}$, $a=0.12$, $b=0.39$, $c=1.39$ \\
Heuristic &  $L_{\gamma} = K \dot E^{a}$ & $K=10^{16.5}$, $a=0.5$ \\
OG & $L_{\gamma} = 10^{47.63} f_{\gamma} P^{\alpha} \dot P^{\beta}$ & $f_{\gamma} = 0.0116$ , $\alpha = -1.93$, $\beta = 0.75$ \\
PSPC & $L_{\gamma} = 10^{47.63} f_{\gamma} P^{\alpha} \dot P^{\beta}$ & $f_{\gamma} = 0.0117$ , $\alpha = -2.43$, $\beta = 0.90$ \\
TPC & $L_{\gamma} = 10^{47.63} f_{\gamma} P^{\alpha} \dot P^{\beta}$ & $f_{\gamma} = 0.0122$ , $\alpha = -2.12$, $\beta = 0.82$2
\enddata 
\end{deluxetable*}

\section{Results}
\label{sec:results}

We evaluate different models based on two main approaches. First, we simply look at the number of Fermi-LAT detectable $\gamma-$ray pulsars predicted by each model and compare it with the known number of pulsars plus the number of PLUIDs in the 4FGL-DR4 Fermi-LAT catalog. Second, we look at the distribution of the modeled sample's period ($P$), period derivative ($\dot{P}$), Galactic longitude, $(l)$ and latitude $(b)$ distributions to see how they compare with the observed sample. We do this as a three step qualitative process by first looking at the distribution in the $P-\dot{P}$ plot, then comparing the parameters' cumulative distribution functions (CDFs) and finally evaluating their quantile-quantile  (QQ) plots \citep[][]{9ba8158e-d8f3-3f5b-8e59-a801be8dc025}. For a given CDF, every probability value $p$ from the ordinate can be mapped to a quantile value $q(p)$ on the abscissa. For a particular parameter, plotting the $q(p)$ function for two different distributions (in this case, observed CDFs versus model CDFs of the parameter) on the same two-dimensional plot results in a QQ plot for that parameter. If both the distributions are identical, the resulting plot would be a straight line with a gradient of one, passing through the origin. More importantly, if either of the two distributions is a linear function of the other, the resulting QQ plot would still be a straight line albeit with a different slope and $y$-intercept. Any differences between the two distributions, specially in their tails, can be easily spotted on a QQ plot as deviations from linearity. These deviations are easier to spot then noticing subtle changes in the CDFs themselves which is why QQ plots were employed as a visual statistical tool in this work. {The purpose of these investigations was not to select an overall preferred model, but rather to check how the underlying distribution resulting from each model compares with the observed one for a given parameter}. We also show the $P-\dot P$ plot for the resulting populations from our radio synthesis compared with the observed sample which was used to run the Monte Carlo simulation in Fig.~\ref{fig:p-pdot-all}.

\subsection{Millisecond pulsars (MSPs)}

\begin{figure}[b]
\plotone{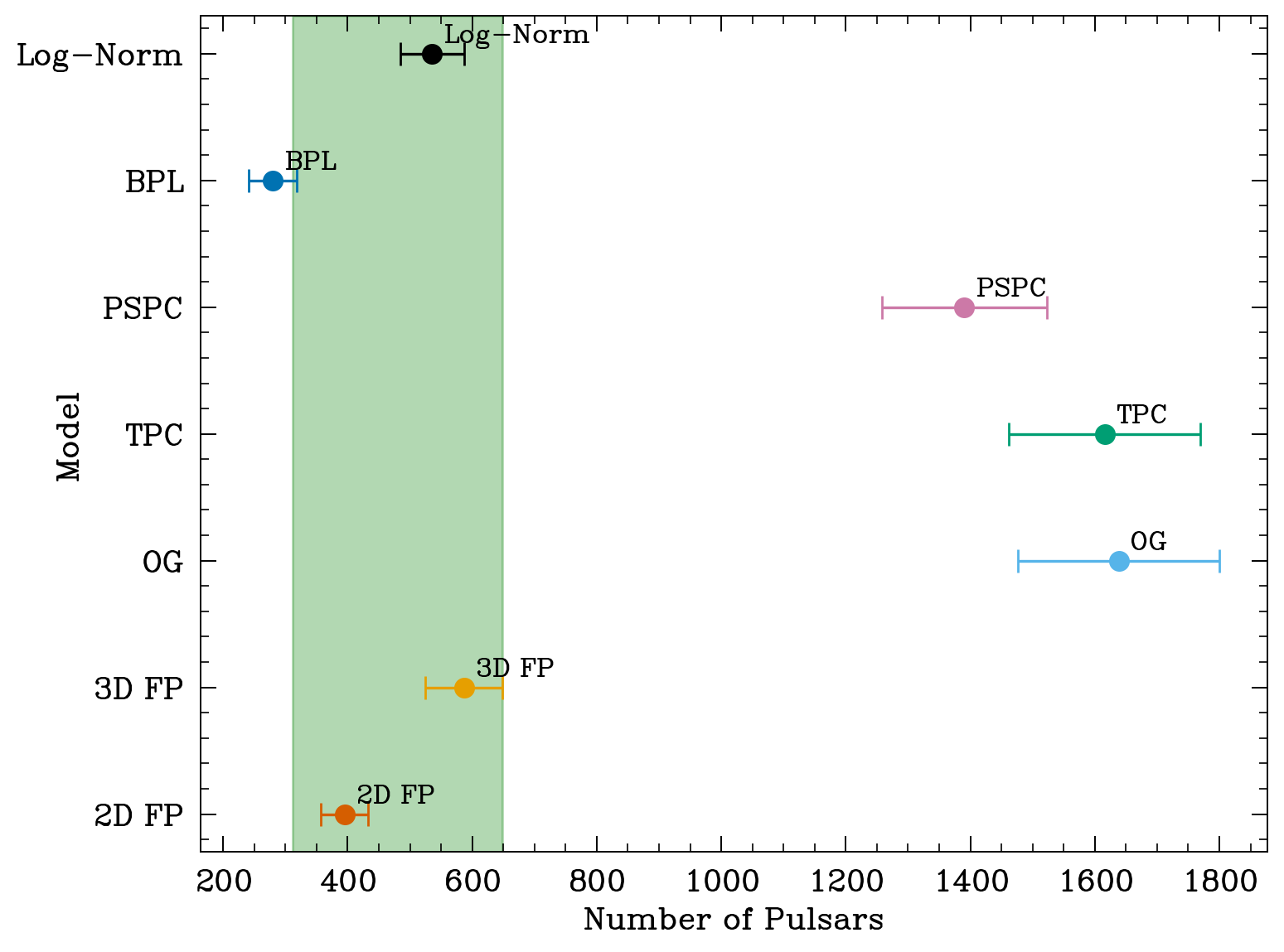}
\caption{The number of Fermi-LAT detectable MSPs for each luminosity model. As described in Section 2, the green highlighted region is the acceptable range (312--648) for any model. 
\label{fig:numbers}}
\end{figure}

\begin{figure}[ht!]
\plotone{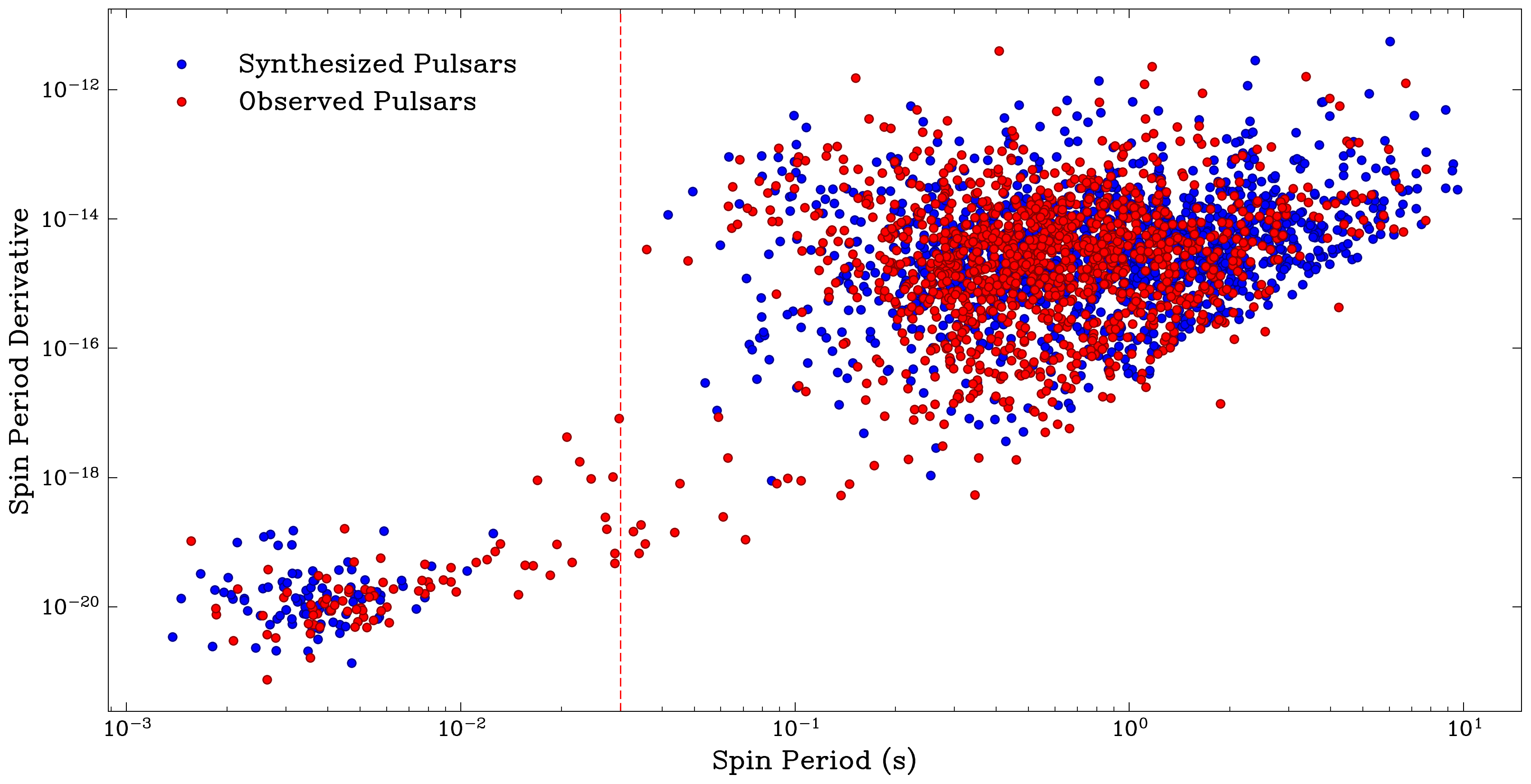}
\caption{ $P-\dot P$ diagram showing the observed and synthesized radio population of CPs and MSPs. The red dashed line at a period of 30 ms separates the MSPs from the CPs. 
\label{fig:p-pdot-all}}
\end{figure}

\begin{figure}[ht!]
\plotone{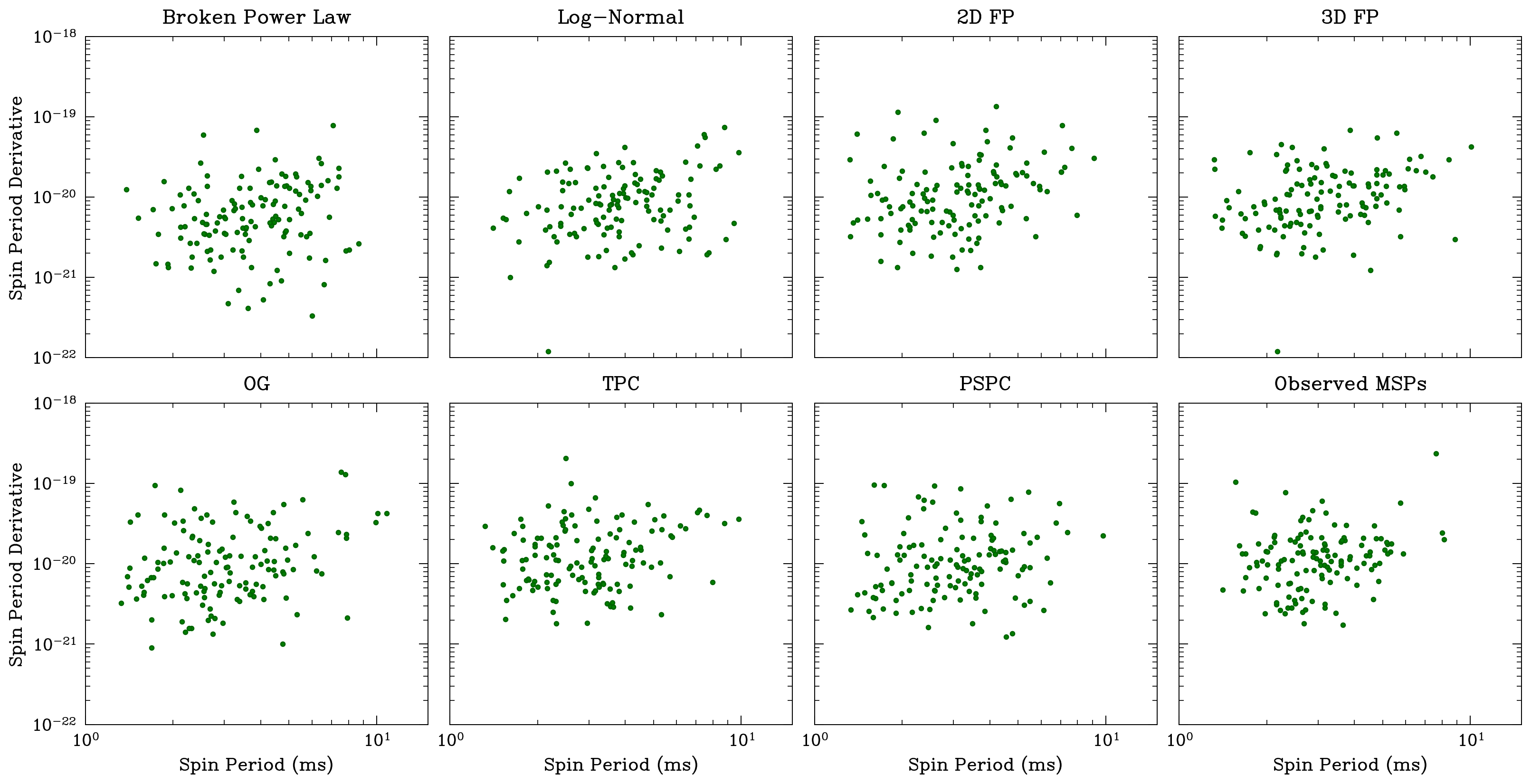}
\caption{Scatter plots showing the $P-\dot P$ distribution for detectable $\gamma-$ray MSPs as predicted by the seven different luminosity models and the observed sample. Each plot contains 134 MSPs.
\label{fig:p-pdot-msps-all}}
\end{figure}

\begin{figure}[ht!]
\plotone{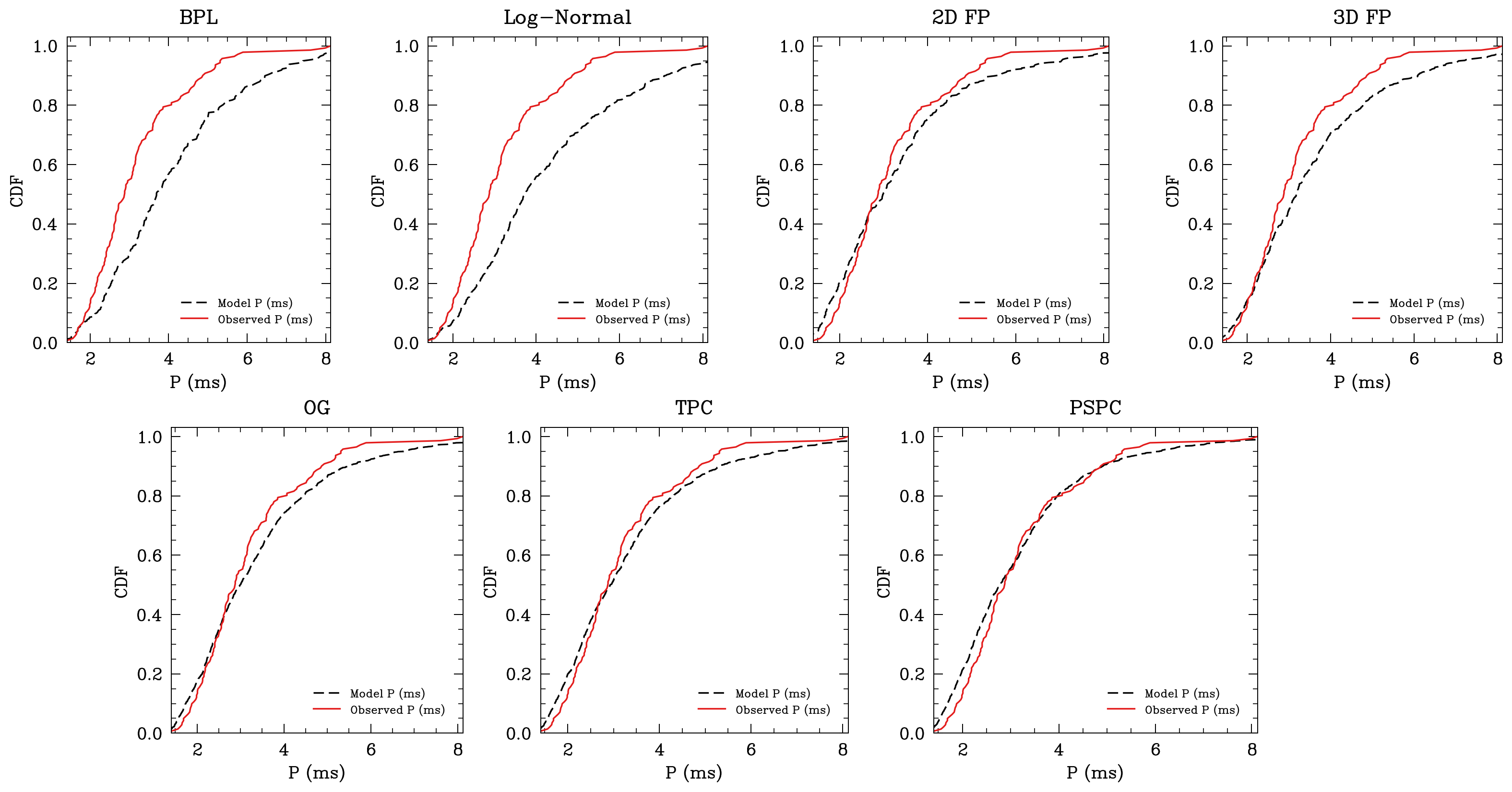}
\caption{CDFs showing the distribution of period for the observed and model populations of MSPs. The red, solid line in each plot represents the observed population and is labeled as ``Obs'' in the legend. The dashed, black line represents the model population and is labeled as ``Mod'' in the legend.
\label{fig:p-hist}}
\end{figure}

\begin{figure}[ht!]
\plotone{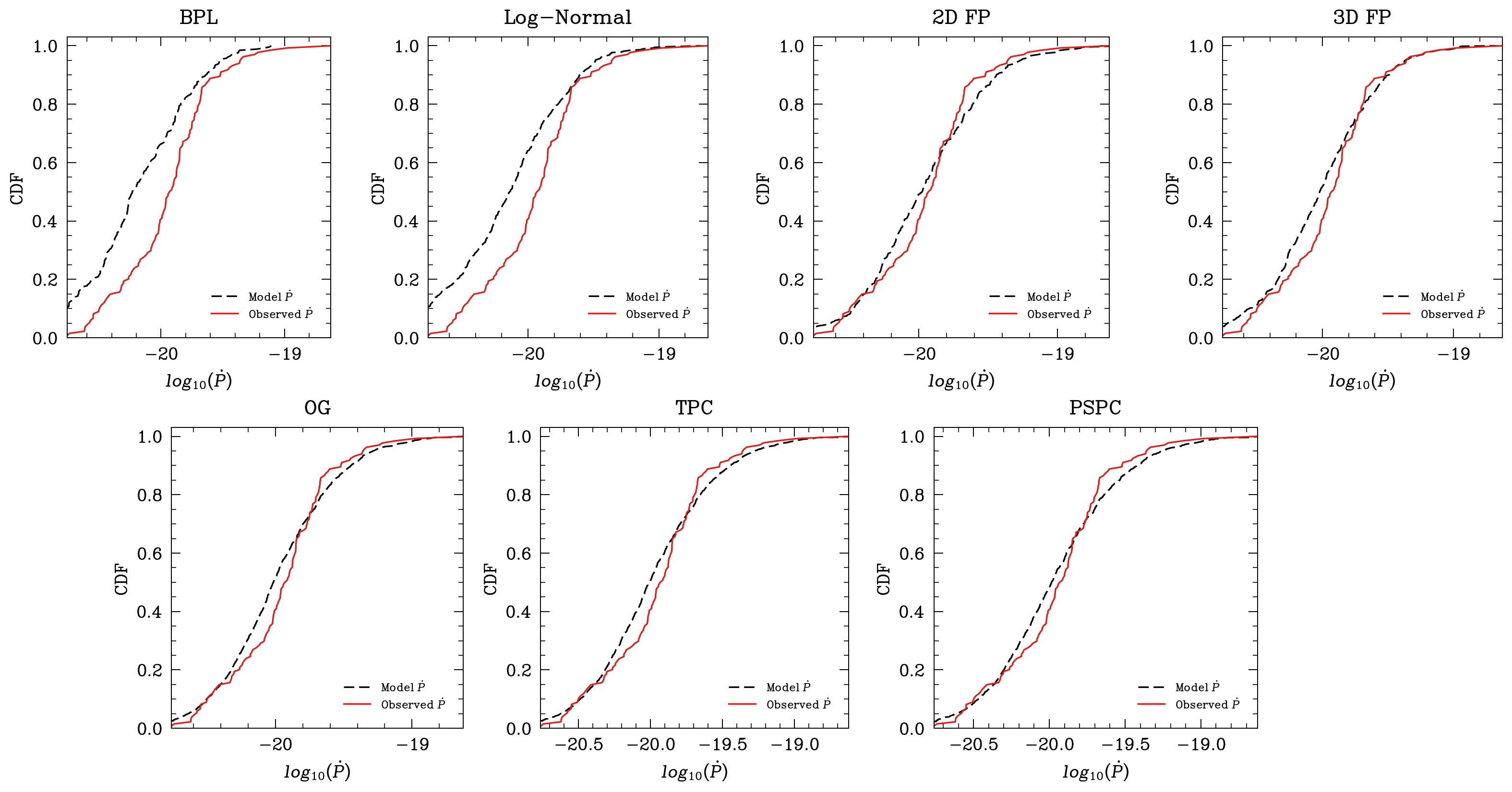}
\caption{CDFs showing the distribution of period derivative for the observed and model populations of MSPs. The red, solid line in each plot represents the observed population and is labeled as ``Obs'' in the legend. The dashed, black line represents the model population and is labeled as ``Mod'' in the legend.
\label{fig:pdot-hist}}
\end{figure}

\begin{figure}[ht!]
\plotone{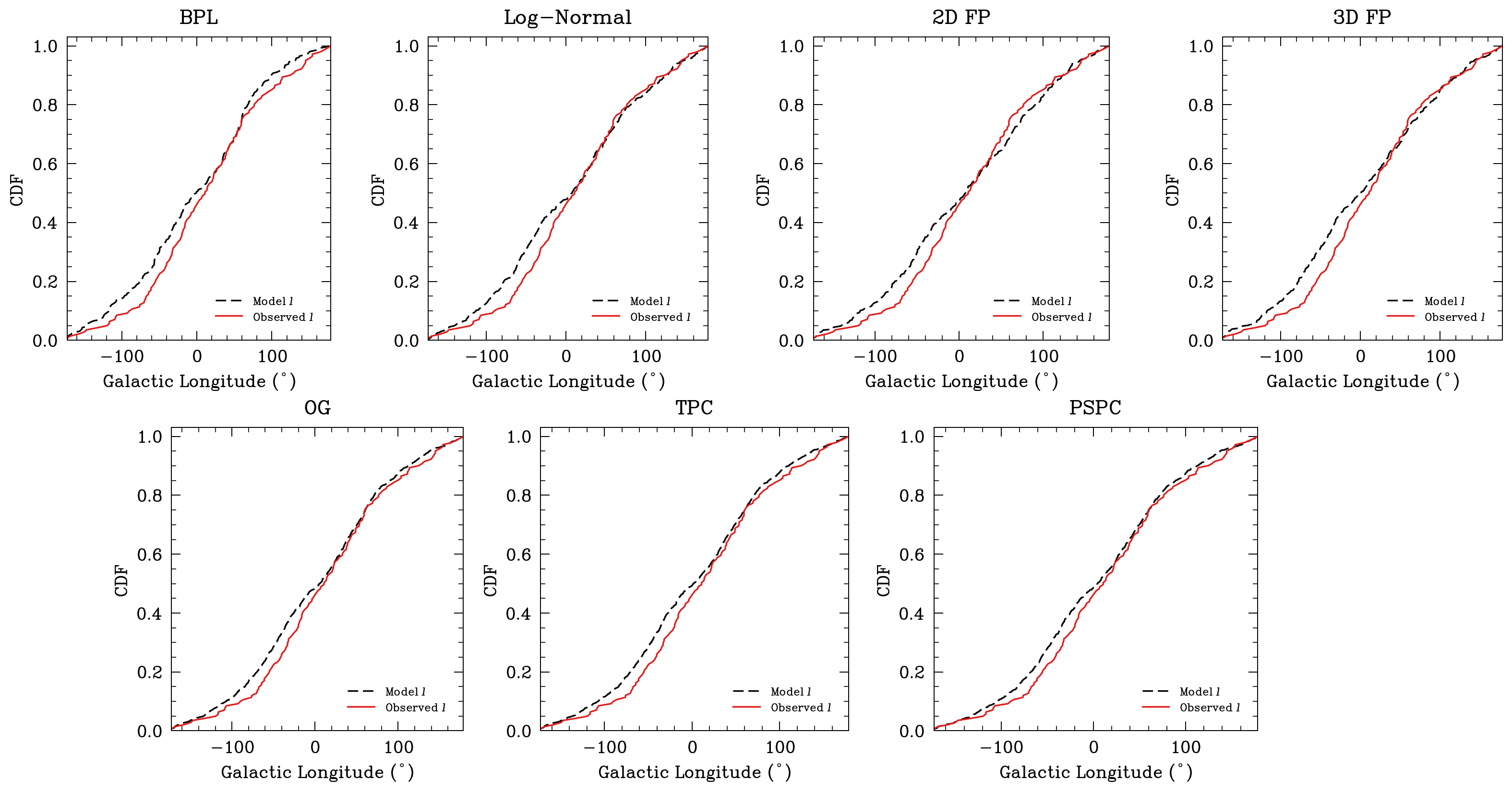}
\caption{CDFs showing the distribution of Galactic longitude for the observed and model populations of MSPs. The red, solid line in each plot represents the observed population and is labeled as ``Obs'' in the legend. The dashed, black line represents the model population and is labeled as ``Mod'' in the legend.
\label{fig:gl-hist}}
\end{figure}

\begin{figure}[ht!]
\plotone{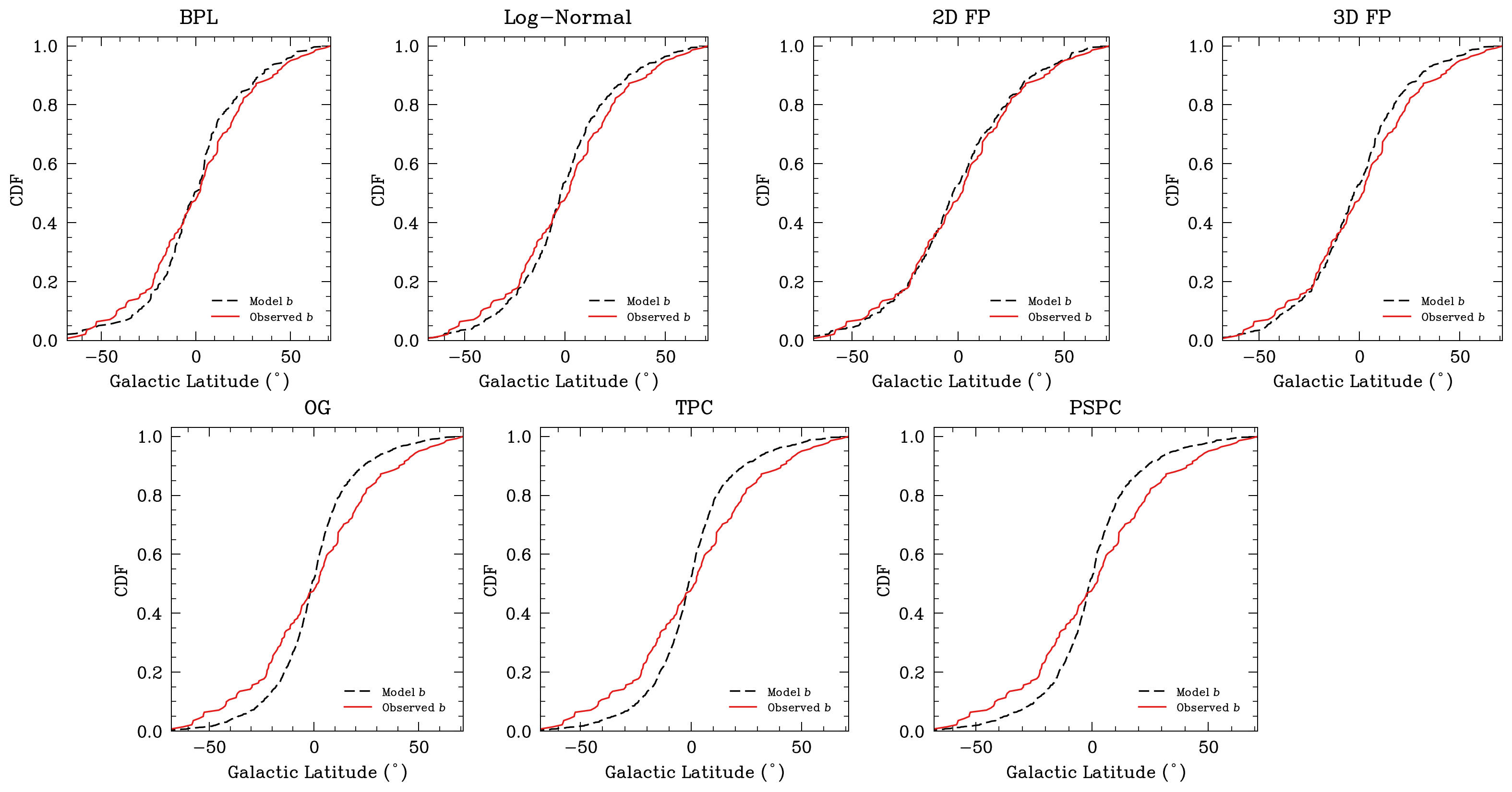}
\caption{CDFs showing the distribution of Galactic latitude for the observed and model populations of MSPs. The red, solid line in each plot represents the observed population and is labeled as ``Obs'' in the legend. The dashed, black line represents the model population and is labeled as ``Mod'' in the legend.
\label{fig:gb-hist}}
\end{figure}

We show the number of Fermi-LAT detectable $\gamma-$ray pulsars predicted by each luminosity model in Fig. \ref{fig:numbers}, along with the acceptable region described in Section~\ref{sec:catalog} highlighted in green. We compare the observed and synthesized distribution for four of these parameters by looking at CDFs shown in Figs.~\ref{fig:p-hist}, \ref{fig:pdot-hist}, \ref{fig:gl-hist} and \ref{fig:gb-hist}. The observed sample contains 141 $\gamma-$ray MSPs which are in the Galactic plane (i.e., excluding globular cluster MSPs J1823$-$3021A and J1824$-$2452A) except for the $\dot P$ CDFs in \ref{fig:pdot-hist}. The observed sample in $\dot P$ CDFs contains 134 MSPs because 7 MSPs do not have a recorded $\dot P$ values in 3PC. Finally, we also provide a $P-\dot P$ plot for each of the models in Fig.~\ref{fig:p-pdot-msps-all}. In this figure, each $P-\dot P$ plot has the same number of pulsars. This was done by randomly sampling 134 pulsars to match the number of known pulsars which have measured $\dot P$ and are shown in the same figure.

\begin{deluxetable}{ccc}[t]
\tablewidth{1.0\linewidth}
\tablecaption{Number of detectable $\gamma-$ray MSPs as predicted by each $\gamma-$ray luminosity model. Radio detectable means the MSP has a radio flux density of at least 100~$\mu$Jy. \label{tab:4}
}
\tablecolumns{3}
\tablehead{
\colhead{$\gamma-$ray luminosity model} & \colhead{$\gamma-$ray detectable} & \colhead{$\gamma-$ray \& Radio detectable}
}
\startdata
BPL & $279 \pm 38$ & $59 \pm 11$ \\
LN & $535 \pm 51$ & $111 \pm 9$ \\
2D FP & $395 \pm 38$ & $134 \pm 12$  \\
3D FP & $587 \pm 62$ & $154 \pm 16$ \\
OG & $1683 \pm 161$ & $252 \pm 17$  \\
PSPC & $1391 \pm 132$ & $234 \pm 17$  \\
TPC & $1616 \pm 153$ & $251 \pm 19$  \\
\enddata
\end{deluxetable}

\subsection{Canonical pulsars (CPs)}

\begin{figure}[ht!]
\plotone{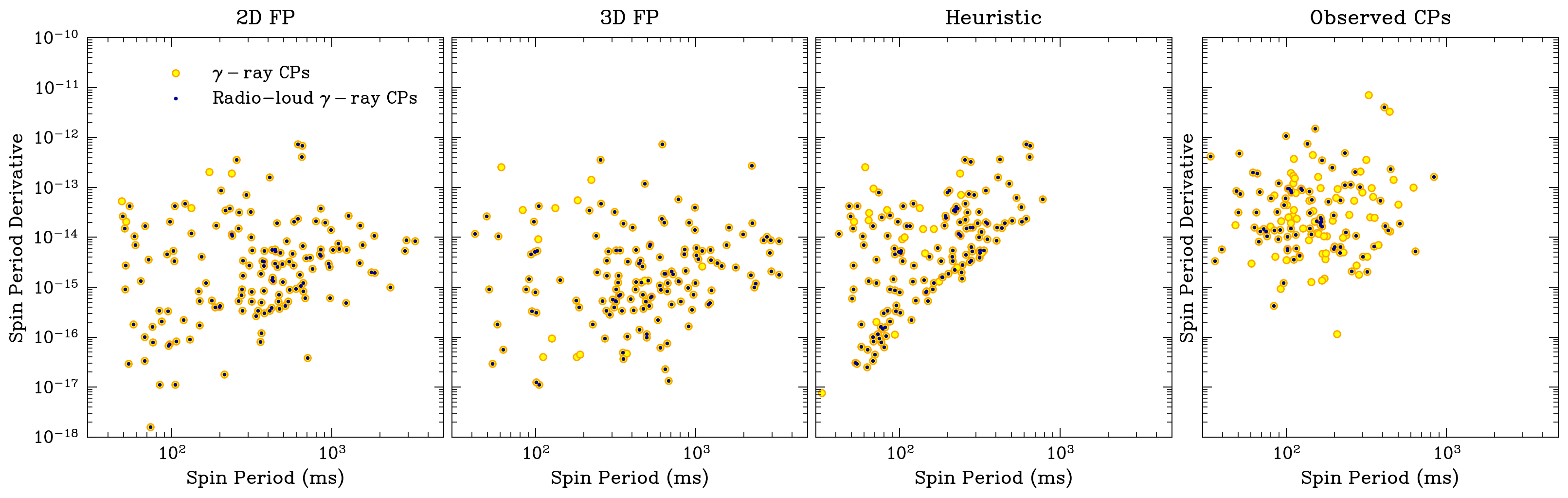}
\caption{$P-\dot P$ distributions for detectable $\gamma-$ray and radio loud $\gamma-$ray CPs for three luminosity models along with the observed sample.}
\label{fig:p-pdot-cps-all}
\end{figure}

The 3D-plane suggests 440 Fermi-LAT detectable $\gamma-$ray sources, the 2D-plane suggest 257 of them and the heuristic model, which included beaming, suggests 157 of them. We show the $P-\dot P$ distribution for CPs compared with the observed sample in Fig. \ref{fig:p-pdot-cps-all}. Since the number of predicted CPs is more than the observed sample for all three models, we chose to show {only} 150 randomly sampled CPs for each model in these plots. We also show the number of radio-loud $\gamma-$ray sources in this plot. We define radio-loud as the CP having a radio flux of greater than  100~$\mu$Jy.


\begin{figure}[ht!]
\plotone{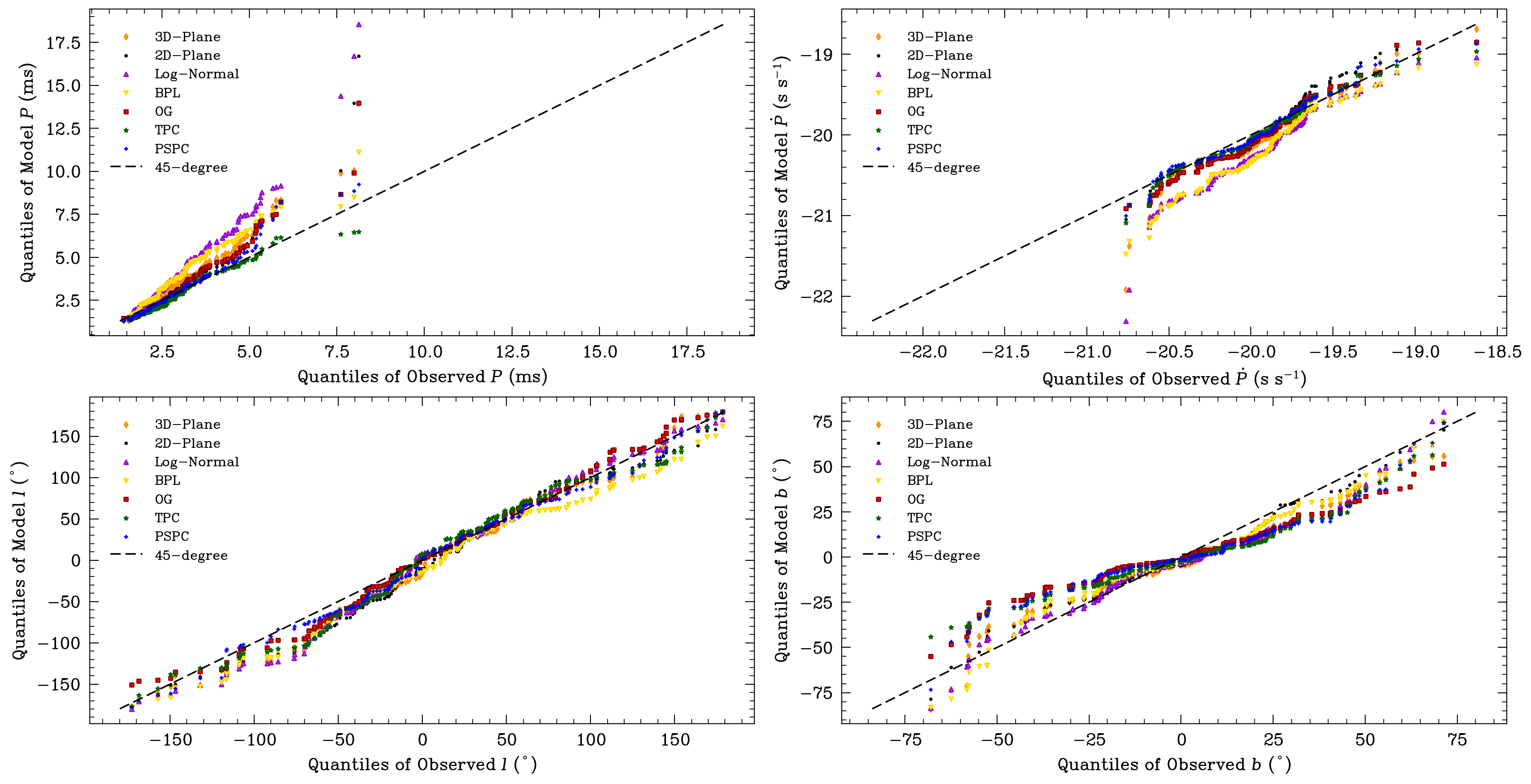}
\caption{QQ plots showing the $P$, $\dot P$, $l$ and $b$ distribution for detectable $\gamma-$ray MSPs as predicted by the different luminosity models compared with the observed sample. $\log(\dot P)$ was used instead of $\dot P$ for the period derivative analysis. All samples used are the same size of 141 except in the $\dot P$ panel. The $\dot P$ panel sample size is 134 due to a lack of $\dot P$ for 7 observed MSPs. 
\label{fig:qq-plot}}
\end{figure}

\section{Discussion}

It is important to assess that the radio synthesis results in a population compatible with the observed radio pulsars. This can be done by qualitatively using the $P- \dot P$ plot shown in Fig.~\ref{fig:p-pdot-all}. We can see that the CP synthesis does well, as expected from the work of \citet{2017MNRAS.467.3493J}. Similarly, the MSP synthesized sample is able to cluster around the same point as the observed sample. The main discrepancy between the model and observed pulsars in Fig.~\ref{fig:p-pdot-all} is the lack of synthesized pulsars in the 20--100~ms period range. This is not surprising given that our modeling does not account for high-mass binary evolution which is thought to produce most of the pulsars in this region
\citep[i.e., the double neutron star binaries and the disrupted recycled pulsars;][]{2023pbse.book.....T}. These systems are not expected to have a significant impact on our main findings.

\subsection{Millisecond pulsars (MSPs)}

As can be seen in Fig.~\ref{fig:numbers}, only the log-normal, 2D-plane and the 3D-plane $\gamma-$ray luminosity model is in the acceptable region for number of Fermi-LAT detectable MSPs in the Galaxy. None of the other $\gamma-$ray luminosity models proposed in the literature so far are compatible with a population synthesis calibrated to the radio population. However, the observed sample size of PLUIDs we used was based on simple assumptions. More complicated methods which could arguably better predict the number of pulsars among this sample also exist in literature. These methods use classification from machine learning (ML) to untangle the nature of the unidentified sources. \citet{2024MNRAS.527.1794Z} use multiple ML models and then create a voting ensemble from those to classify unidentified sources (UIDs) after dividing the sample into a high Galactic latitude (HGL) region where $|b| > 10^{\circ}$ and a low Galactic latitude (LGL) region where $|b| < 10^{\circ}$. Their ensemble finds 88 PLUIDs in the HGL and 134 in the LGL regions respectively. The fact that they are unable to distinguish between MSPs and CPs makes a direct comparison with our work difficult. However, the vast majority of the HGL region pulsars are likely to be MSPs and the LGL region should have a CP majority. {This assumption is supported by looking at the percentage of $\gamma-$ray MSPs and CPs that are not in the HGL and LGL region respectively. 28$\%$ of MSPs are not in the HGL region and only 10$\%$ of CPs are not in the LGL region. To get a lower limit, we can assume that the percentage of PLUIDs in the HGL region that can be MSPs is the same as the percentage of observed $\gamma-$ray MSPs that are in HGL region. This suggests a lower limit of $(0.72 \times 88)+134=197$ MSPs to be compatible with their work.} All of our models satisfy this limit. {Similarly, we can get an upper limit assuming all the HGL PLUIDs are MSPs and 28$\%$ of LGL PLUIDs are MSPs as well. This gives us an upper limit of $88+(0.28\times134)+134=260$. None of our models satisfy this upper limit. This suggests that either the ML classifier is overfit or all the luminosity models are overestimating $\gamma-$ray luminosities. The latter could be true because the current sample of $\gamma-$ray MSPs is biased towards MSPs with the highest $\gamma-$ray luminosities. It would also imply that the $\gamma-$ray luminosity modeling so far has not been able to correctly account for this bias.}

Looking at the $P-\dot P$ distributions in Fig.~\ref{fig:p-pdot-msps-all}, we see that the observed sample clusters around the point at 3~ms and $10^{-20}$, respectively. This is not the case for the BPL and log-normal model as those pulsars have {a wider spread in} $\dot P$s. The CDFs for these parameters help us understand the spread in these $P-\dot P$ plots in a {better} way. In Fig.~\ref{fig:p-hist}, we see that all models except BPL and Log-normal match the observed sample until {a} period of 3~ms. 3D FP and OG model lack the required density in the 3--5~ms period interval. TPC and 2D FP are able to {better} match the observed sample in {that} range but {are} underdense in the 5--10~ms range. PSPC matches the observed distribution the closest. Similarly, the $\dot P$ distribution in Fig.~\ref{fig:pdot-hist} shows us that the BPL and log-normal models perform the worst. Overdensity in the $\dot P < 10^{-{19.8}}$~s~s$^{-1}$ {region} is seen for all models. {All models do better in the higher $\dot P$ range, with 3D FP being able to best match the observed sample}. The sky distributions for these simulated populations is shown in Figs. \ref{fig:gl-hist} ($l$) and \ref{fig:gb-hist} ($b$). The $l$ distributions are all in a reasonable agreement with the observed sample. Their noticeable features are that all models have a larger population under $l < 0$ than the observed. The trend continues for $l > 0$ for all models except the for 2D and 3D FP. The $b$ CDFs for all models except the GHF+18 models is also in close agreement with the observed distribution. The GHF+18 models are closer to the galactic plane than the observed sample.

We can find further insights into these distributions by looking at how each quantile of the parameter distributions compares with each other in a QQ plot. For MSPs, this is shown in Fig.~\ref{fig:qq-plot}. All $\gamma-$ray luminosity models {except for OG produce high period ($P > 9$~ms) MSPs which do not exist in the observed population. Between the $3-5$~ms range, we see a discrepancy between the models. Here, the BPL, LN and 3D FP are more sparse than the observed sample and other models. The $\dot P$ distributions in Fig.~\ref{fig:qq-plot} show that BPL, LN and 3D FP models result in a distribution with longer tails than the observed distribution. All models except BPL and LN do well for rest of the range of $\dot P$s until the upper tail. At the upper end, only OG is able to match the observed distribution.} The $l$ QQ plot is the least interesting since it shows that all the models are in reasonable agreement with the observations since they all lie close to the diagonal line. On the other hand, the $b$ QQ plot has an S-shaped distribution which clearly indicates that the model distributions have heavy tails. If any of these models is assumed to be an acceptable representation of the real sample, {then} these heavy tails can be interpreted as showing the bias towards lower latitudes in the observed sample. This would mean that there are more MSPs waiting to be discovered further away from the Galactic plane.

\begin{figure}[ht!]
\plotone{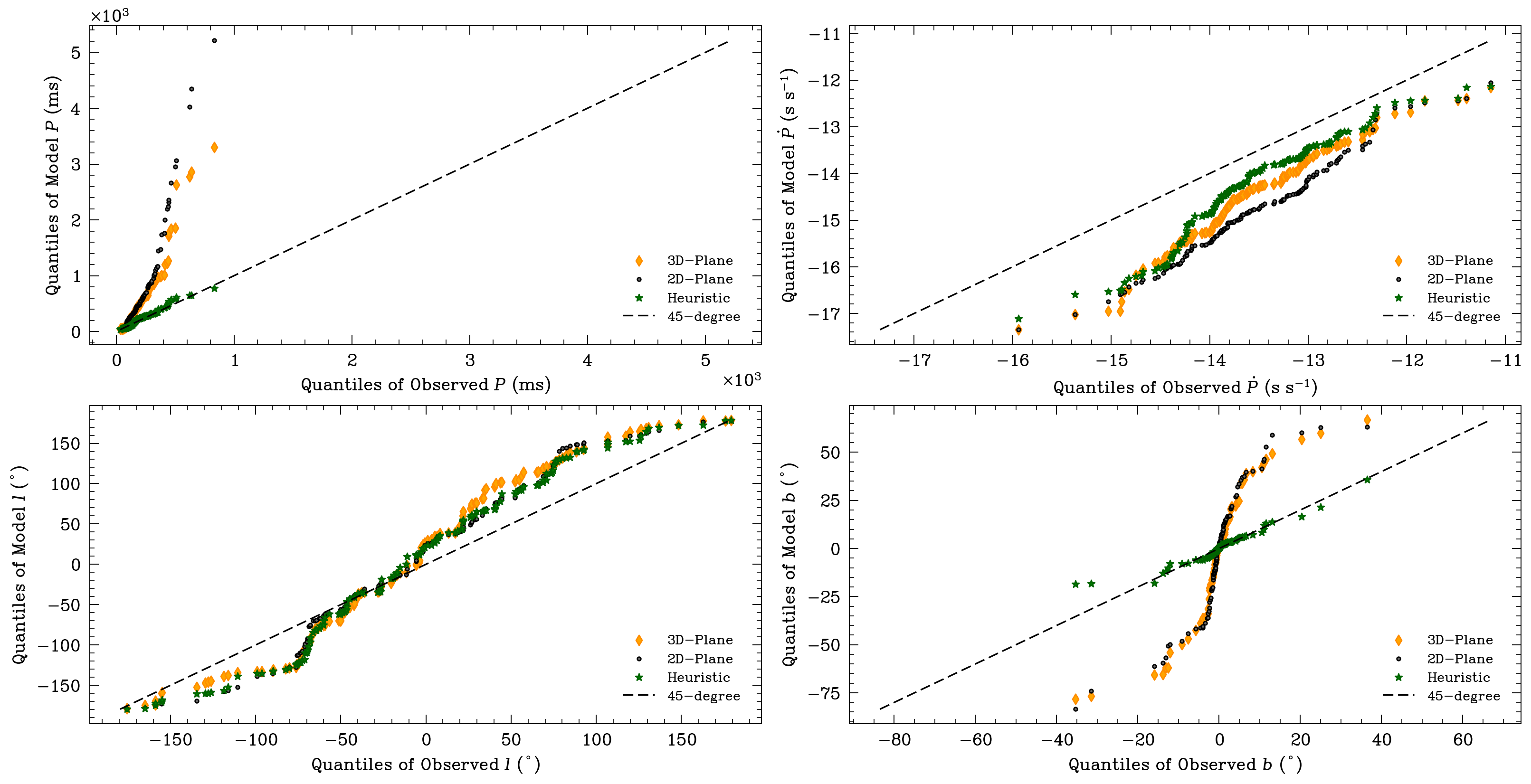}
\caption{QQ plots showing the $P$, $\dot P$, $l$ and $b$ distribution for detectable $\gamma-$ray CPs as predicted by the two FP luminosity models compared with the observed sample. All samples used are the same size of 150. $\log(\dot P)$ was used instead of $\dot P$ for the period derivative analysis. \label{fig:qq-cp}}
\end{figure}

\subsection{\label{sec:cp-discussion}Canonical pulsars (CPs)}

 All three of the luminosity models used for CPs result in a number of predicted $\gamma-$ray pulsars compatible with the lower limit of 150 and the upper limit of 705 that we derived in Section \ref{sec:catalog}. Looking at their distribution in the $P-\dot P$ plane shown in Fig. \ref{fig:p-pdot-cps-all}, we see that the model distributions are not compatible with the observed sample. Both of the FP models contain an excess of low $\dot P$ and high $P$ CPs. For the heuristic model, we initially tried an $\dot{E}$ dependent scheme as proposed by \cite{2024arXiv240611428S}. However, this also produced an excess of low $\dot P$ and high $P$ CPs compared to what is observed. The only way we could get any type of resemblance was to, arbitrarily, adopt a very sharp cutoff in the beaming in which the fraction is unity if $\dot{E}>10^{32}$erg~s$^{-1}$ and zero otherwise. Even this ad hoc approach, produces too many low $\dot P$ CPs.  

We show the QQ-plots for CPs in Fig.~\ref{fig:qq-cp}. Looking at the period, we see that the FP models result in a very wide distribution extending to 3~s whereas the maximum period in the observed sample is 0.83~s. As was already seen in the $P-\dot P$ plot, the heuristic model results in a similar distribution to the observed sample. The $\dot{P}$ panel shows that the models consistently under-predict the number of CPs at high $\dot P$ while also over-predicting low $\dot P$ CPs. Here as well, the heuristic model is the closest to the observed distribution. The $l$ distribution shows an interesting feature at around $-70^{\circ}$ where we see a sudden jump suggesting a gap in the model sample densities. The $b$ distribution shows that the FP models result in distributions with wider tails than is seen in the observed distribution. There are far more CPs away from the Galactic plane in the FP model samples than are observed. The heuristic model is in clear contrast with the FP models by falling on the reference line; an indication of it matching the observed distribution. We investigated the shortfall of the FP models spatial distribution further. The $P$ and $\dot P$ distributions from these models can help us understand it. We see that these models are predicting too many low-$\dot P$ and long-period CPs. This indicates that our synthesized sample has an older population of CPs than is seen in the observed sample. As a result, the older population has also moved further away from the Galactic plane compared to the observed sample. We test this hypothesis by capping the maximum characteristic age, defined as $\tau = P/2 \dot P$, of our synthesized 3D FP and 2D FP CPs to the maximum age seen in the observed sample, 3~Myr. After doing this, we see that we are left with a population of CPs which is similar in its  $b$ distribution to the observed sample. We note the caveat that $\tau$ is not a perfect proxy for true age since it assumes a braking index of 3 and a negligible birth period.

\begin{figure}[ht!]
\plotone{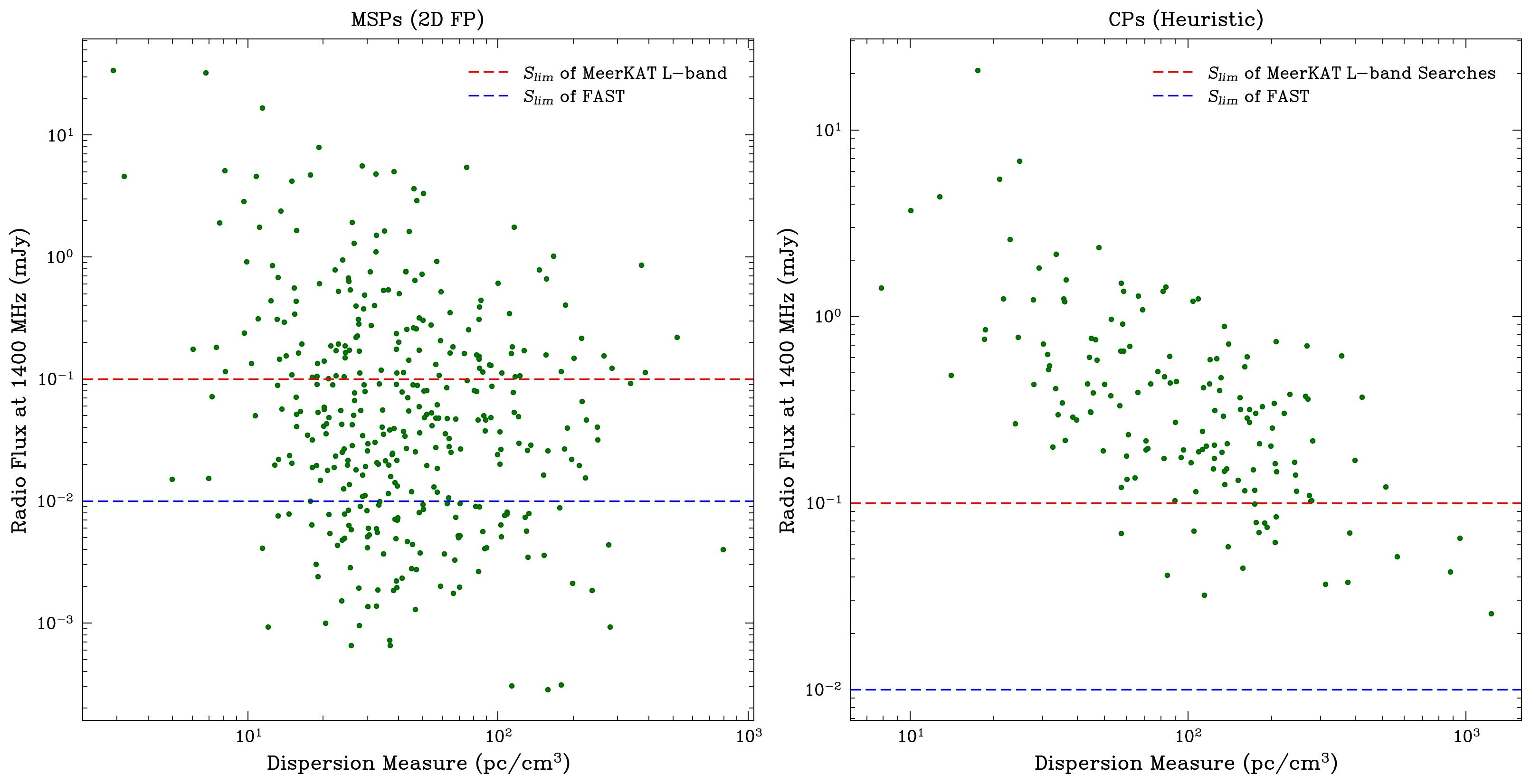}
\caption{Flux vs DM distribution for synthesized MSPs with 2D-FP $\gamma-$ray luminosity model and the heuristic luminosity model for CPs. The red dotted line corresponds to the 100~$\mu$Jy sensitivity limit achieved by targeted radio searches at L-band by TRAPUM using MeerKAT. {The blue dotted line at 10~$\mu$Jy refers to the approximate sensitivity limit of FAST.}
\label{fig:flux-dm}}
\end{figure}

\begin{figure}[ht!]
\plotone{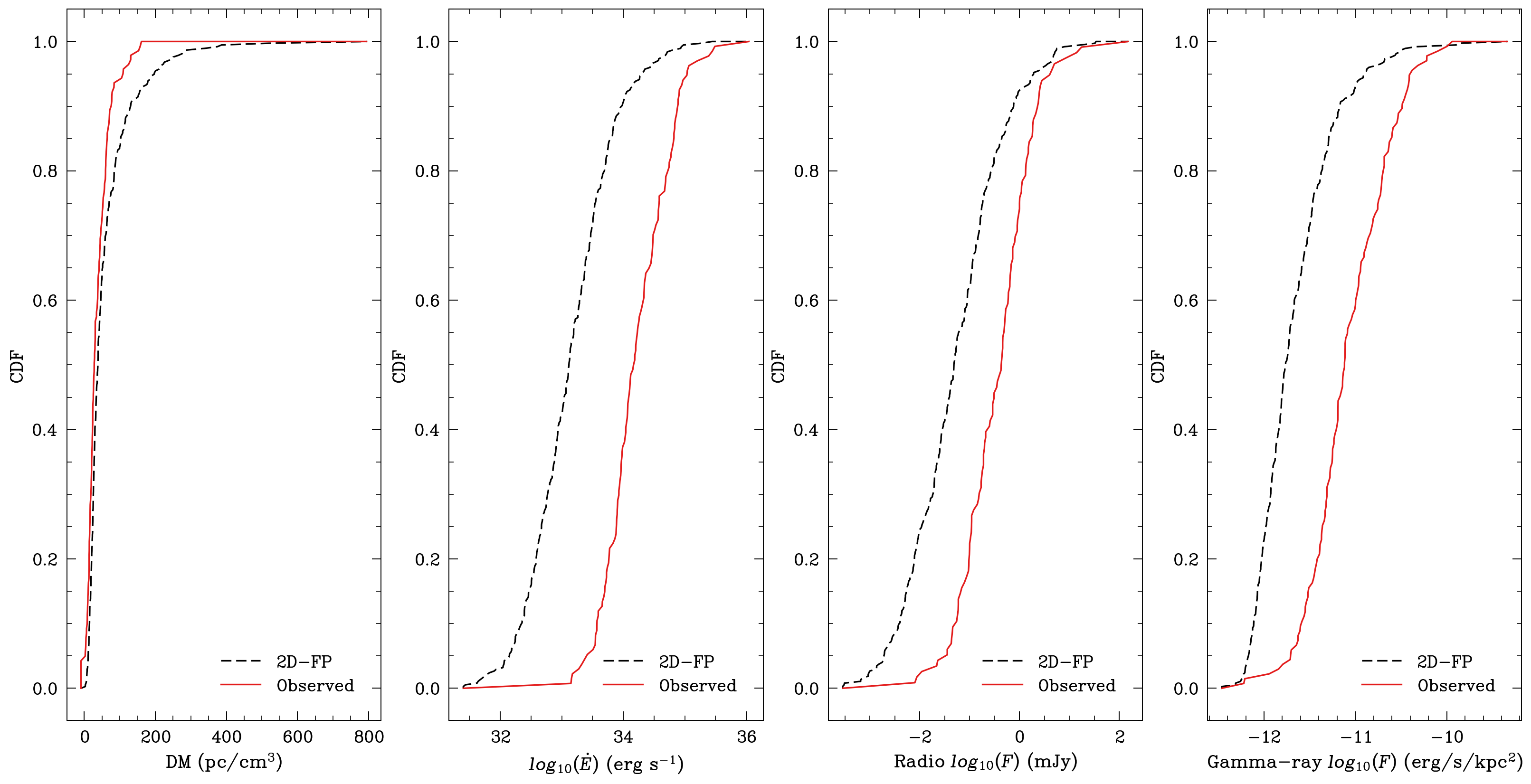}
\caption{CDF of observable quantities for synthesized MSPs with the 2D-FP $\gamma-$ray Luminosity model compared with the 141 observed $\gamma-$ray MSPs. The y-axis is normalized to allow better comparison of the distributions between the observed and synthesized sample. In the observed sample, 14 are missing a radio flux and 7 are missing $\gamma-$ray flux and $\dot E$ values so they are not included in their respective CDFs.
\label{fig:hists-msp}}
\end{figure}

\begin{figure}[ht!]
\plotone{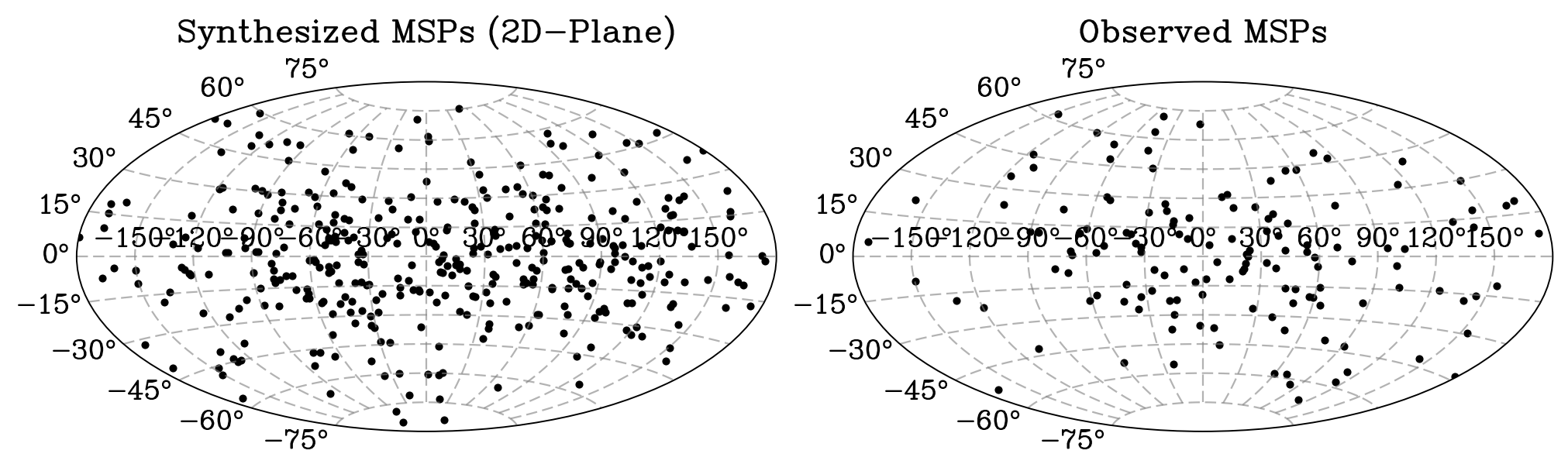}
\plotone{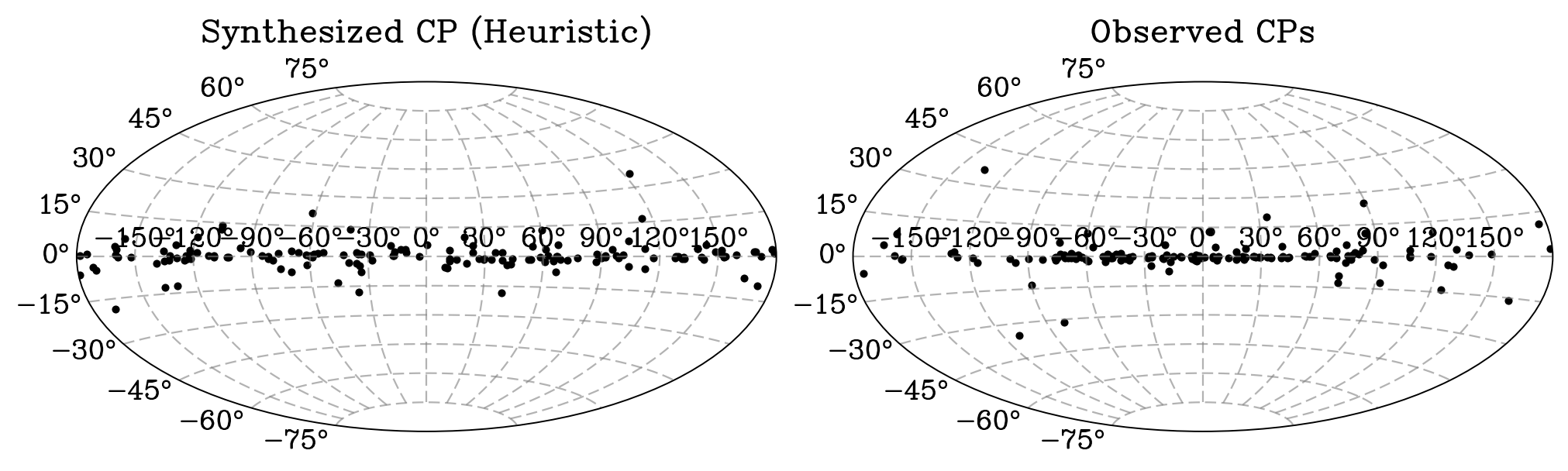}
\caption{Distribution in Galactic coordinates showing known and simulated MSP and CP populations of $\gamma-$ray pulsars.
\label{fig:aitoff-msp}}
\end{figure}

\subsection{Predictions for future radio surveys of PLUIDs}

The synthesized samples for both MSPs and CPs can be used to make predictions for future targeted radio searches of PLUIDs by looking at sources which have radio fluxes above the sensitivity limit of the deepest searches done so far. We use the sensitivity limit from the searches done by \citet[]{2023MNRAS.519.5590C} using MeerKAT, which is 100~$\mu$Jy. {We also use the sensitivity limit of the Five hundred meter Aperture Spherical Telescope (FAST), which is $\sim$10~$\mu$Jy at L-band for a 5~min observation \cite{2025RAA....25a4001H}. To make these predictions, we select a fiducial model as well. For MSPs, the main constraint that we place for this selection is that the model should be within the green region of Fig. \ref{fig:numbers}. From the three models that satisfy this requirement, we make a decision based on the qualitative analysis of the $P-\dot P$ distribution to use 2D FP as our fiducial MSP model.}  In {the MSP panel of} Fig.~\ref{fig:flux-dm}, we see that most of the Fermi-LAT detectable MSPs fall below {the MeerKAT sensitivity limit} There are only 143 sources which are above {the MeerKAT} limit, slightly more than the 138 known radio-loud $\gamma-$ray MSPs. This suggests that, according to {the 2D FP} model, most of the Fermi-LAT detectable MSPs that are loud enough to be detected by {the most sensitive targeted searches that have been done so far} have been discovered. { There are 283 MSPs above the FAST sensitivity limit. This suggests that targeted searches of PLUIDs by FAST could potentially result in discovery of $283-143=140$ MSPs, doubling the current sample size. Discovering more MSPs among the PLUIDs would require probing even lower radio fluxes. This can be achieved by longer observations using existing instruments or by using more sensitive instruments such as the upcoming Square Kilometer Array telescope.}  For CPs, we pick the heuristic model as our fiducial model {due to the discussion about the FP models in section \ref{sec:cp-discussion}. Looking at the DM-flux distribution in Fig.~\ref{fig:flux-dm}, we} see that most of the CPs (137/157) lie above the sensitivity limit {of MeerKAT. All of them are above the limit of FAST.} The number of known radio-loud CPs is just 84. This suggests, if the heuristic model is assumed to be a good approximation of $\gamma-$ray luminosity, there are many more radio CPs waiting to be discovered among PLUIDs using existing instruments. This strongly supports future targeted observations of LGL region PLUIDs.

For the MSPs, to further preview the properties of these future discoveries, we show CDFs for the distribution of DM, $\dot E$, radio flux at 1400 MHz, and $\gamma-$ray flux for {the} 2D-FP model overlaid with the currently observed MSPs in Fig.~\ref{fig:hists-msp}. For DM, we see that the model has a longer tail with DMs up to $\sim$400~pc~cm$^{-3}$ whereas the maximum DM in the observed sample is 161~pc~cm$^{-3}$. This is indicative of the bias towards MSPs closer to Earth, given DM is a proxy for distance. As can be seen in the sky distribution in Fig.~\ref{fig:aitoff-msp}, {the 2D FP model suggests that} there are still a lot of MSPs to be discovered among the PLUIDs in the HGL region. This is especially important for targeted radio searches since pulsars away from the Galactic plane are not only easier to detect due to mitigated levels of scintillation and scattering but also easier to observe since time on radio telescopes is generally easier to obtain due to such regions being away from the highly oversubscribed sidereal time ranges of most observatories. The remaining three distributions shown in Fig.~\ref{fig:hists-msp} for the spin-down energy ($\dot E$), radio flux and $\gamma-$ray flux peak at lower values than the observed sample showing that the observed sample is biased towards high $\dot{E}$ sources and/or those closer to Earth.

\section{Conclusions}

We investigated pulsar $\gamma-$ray emission mechanisms proposed by \cite{2018MNRAS.481.3966B}, \citet{2018ApJ...863..199G} and \citet{2023ApJ...954..204K} using a population synthesis framework based on Monte Carlo simulations. Utilizing the larger sample size of radio pulsars and assuming that all $\gamma-$ray pulsars emit in radio as well, we started by synthesizing a population which is calibrated to the radio population. The pulsars were seeded in a model galaxy and allowed to evolve in a model of the Milky Way’s gravitational potential as they spin down before their radio flux is calculated. We calculated $\gamma-$ray luminosities for the populations synthesized by this process and then applied a detection threshold based on the Fermi-LAT sensitivity map. The resulting populations were then compared with the known sample of $\gamma-$ray pulsars. Primary attention was given to the size of the sample predicted by each model and the distributions of period, period derivative, Galactic longitude and latitude. We chose not to present the Kolmogorov-Smirnov or Anderson-Darling tests, to compare the modeled samples with the observed sample. Our purpose is to compare the models and understand the features of the observed sample that the models are able to capture and the features that they are unable to capture. We then analyze how these features change the sample size of detectable $\gamma-$ray pulsars with Fermi-LAT. Future work will hopefully be able to address these features to improve the models and help us better understand the Galactic population of $\gamma-$ray pulsars. 

For MSPs, we found that only the log-normal, 2D FP, and the 3D FP result in a sample of detectable maps that is in the acceptable range given the observed sample. It is to be noted that we can change the normalizing constants in the models of GHF+18 to bring down their number of predicted MSPs. The larger sample size of MSPs since that work provides incentive to rerun the Markov-Chain Monte Carlo analysis of GHF+18 which we predict would result in smaller normalization constants. An alternative interpretation is that the $\gamma-$ray beaming fraction for these models is closer to 25--50\% rather than the 100\% assumed here. Our results do not pick a preferred model but they can be used as evidence against the models failing to predict detectable MSPs within the 308--650 range. They show the different ways in which the modeled populations differ from the observed sample. 

For CPs, all the models tested satisfy the number constraints. However, the two models adopted from the literature produce an inadequate resemblance to the observed sample when distributed in the $P-\dot P$ plane. This trend continues when looking at the distribution of other observables as shown in Fig.~\ref{fig:qq-cp}. In order to get any sort of agreement in $P-\dot{P}$ space, we had to invoke an abitrary cut-off in the $\gamma-$ray beaming fraction to eliminate model pulsars with $\dot{E}<10^{32}$erg~s$^{-1}$. This is a surprising result, especially since, as can be seen in Fig.~\ref{fig:p-pdot-all}, the radio synthesis generally does an excellent job at reproducing the observed radio population. Therefore, more work is {needed} to understand the $\gamma-$ray emission in CPs. Future work focused on improving the FP models would also benefit from accounting for larger than the observed sample ages seen in the synthesized sample {of CPs}. Our work also shows that the $\gamma-$ray beaming model of \citet[][]{2024arXiv240611428S} is incompatible with the FP relations since the majority of CPs produced by them have $\dot E$s too low to be affected by beaming. 

Our work demonstrates that many of the PLUIDS shown in Fig.~\ref{fig:aitoff-uids} are consistent with as-yet unidentified radio and $\gamma-$ray pulsars, highlighting the need for further observations of PLUIDs using radio telescopes. Re-analyzing archival observations of PLUIDS using different pulsar searching methods can supplement this effort, specifically for longer period CPs \citep{2020MNRAS.497.4654M}. A larger sample size of $\gamma-$ray pulsars would be beneficial for a number a of scientific cases. These include better modeling of pulsar emission mechanisms, improving our understating of the Galactic $\gamma-$ray budget, potentially discovering unique systems with exceptional scientific value such as a double-pulsar system, increasing the sample size of MSPs for pulsar timing array science \citep[see, e.g.,][]{2010CQGra..27h4013H} and shedding light on compact object evolution \citep[see, e.g.,][]{2023pbse.book.....T}.

\begin{acknowledgments}
\nolinenumbers
We thank David Smith and Simon Johnston for insightful discussions and for sharing code to carry out the $\gamma-$ray flux selection (Smith) and the radio pulsar $P-\dot{P}$ modeling (Johnston). {We also thank the anonymous referee for their valuable feedback which helped improve this manuscript.} We acknowledge support from a graduate stipend provided by the Eberly College of Arts and Sciences at West Virginia University.
\end{acknowledgments}

\bibliography{paper}{}
\bibliographystyle{aasjournal}

\end{document}